\documentclass[aps,prd,showpacs,twocolumn,floatfix,nofootinbib,superscriptaddress,
amsmath,amssymb,amsfonts,]{revtex4-1} 

\usepackage{bm}
\usepackage{slashed}
\usepackage{graphicx}
\usepackage{subfigure}
\usepackage[svgnames]{xcolor}

\usepackage{placeins}

\usepackage{xspace}

\def\LEPT{\texttt{LEPTO}\xspace} 
\def\MLEPT{\texttt{\lowercase{m}LEPTO}\xspace} 

\def\vfR{$\varphi_R$\xspace}
\def\fR{\varphi_R}
\def\fS{\varphi_S}
\def\vfT{$\varphi_T$\xspace}
\def\fT{\varphi_T}
\def\kT{\vect{k}_T}
\def\kt{k_T}

\def\Pp#1{\vect{P}_{#1\perp}}
\def\PT#1{\vect{P}_{#1T}}
\newcommand{\vT}{\vect{P}_T}
\newcommand{\vect}[1]{\boldsymbol{#1}}
\newcommand{\bfk}{\vect{k}}
\newcommand{\bfq}{\vect{q}}

\newcommand{\bfS}{\vect{S}_{_T}}

\newcommand{\al}[1]{\begin{align} #1 \end{align}}
\newcommand{\non}{\nonumber}
\newcommand{\vf}{\varphi}

\newcommand{\Ge}{\mathrm{GeV}}
\newcommand{\Gs}{\mathrm{GeV}^2}
\newcommand{\ImS}{0.8\columnwidth}
\newcommand{\Eq}[1]{Eq.~(\ref{#1})}

\begin{document}

\title{Sivers effect in dihadron semi-inclusive deep inelastic scattering}

\preprint{ADP-14-15/T873}

\author{Aram~Kotzinian}
\affiliation{Yerevan Physics Institute,
2 Alikhanyan Brothers St.,
375036 Yerevan, Armenia
}
\affiliation{INFN, Sezione di Torino, 10125 Torino, Italy
}

\author{Hrayr~H.~Matevosyan}
\affiliation{ARC Centre of Excellence for Particle Physics at the Tera-scale,\\ 
and CSSM, School of Chemistry and Physics, \\
The University of Adelaide, Adelaide SA 5005, Australia
\\ http://www.physics.adelaide.edu.au/cssm
}

\author{Anthony~W.~Thomas}
\affiliation{ARC Centre of Excellence for Particle Physics at the Tera-scale,\\     
and CSSM, School of Chemistry and Physics, \\
The University of Adelaide, Adelaide SA 5005, Australia
\\ http://www.physics.adelaide.edu.au/cssm
}

\begin{abstract}
The Sivers effect describes the correlation of the unpolarized parton's transverse momentum with the transverse spin of the nucleon. It manifests as a sine modulation of the cross section for single-hadron semi-inclusive deep inelastic scattering (SIDIS) on a transversely polarized nucleon with the azimuthal angle between the produced hadron's transverse momentum and the nucleon spin ($\varphi_h$ and $\fS$, respectively). It has been recently suggested that the Sivers effect can also be measured in two-hadron SIDIS process as sine modulations  involving the azimuthal angles \vfT~and \vfR~of both the total and the relative transverse momenta of the hadron pair. Here we present the detailed derivation of the two-hadron SIDIS cross section using simple parton-model inspired functional forms for both the parton distribution and the fragmentation functions. We show explicitly that the terms corresponding to the $\sin(\fR-\fS)$ and $\sin(\fT-\fS)$ modulations are nonzero. Further, we derive the cross section expressions  for single-hadron production in the two-hadron sample. Finally, we employ a modified version of the \LEPT Monte Carlo event generator that includes the Sivers effect to estimate the size of single spin asymmetries corresponding to these modulations independent of the formalism developed.
\end{abstract}


\date{\today}                                           

\maketitle

\section{Introduction}
\label{SEC_INTRO}

 The Sivers effect~\cite{Sivers:1989cc} describes the correlation of the transverse momenta of the unpolarized partons with the transverse spin of the nucleon. It is quantified by the Sivers parton distribution function (PDF) and plays an important role in our understanding of the structure of the nucleon. The Sivers PDF can be measured experimentally in single-hadron production in a polarized semi-inclusive deep inelastic scattering (SIDIS) process~\cite{Collins:2002kn}, by measuring the single spin asymmetry (SSA) modulated with respect to the sine of the difference of the azimuthal angle of the produced hadron and the nucleon's transverse spin vector~\cite{Anselmino:2005nn}. Here the SSA is a ratio of a convolution of the Sivers PDF with the unpolarized fragmentation function (FF) to that of the unpolarized PDF and FF. These asymmetries have been measured in HERMES~\cite{Airapetian:2009ae}, COMPASS~\cite{Adolph:2012sp} and, JLab HALL A~\cite{Qian:2011py} experiments.  Nevertheless, due to the  limited range of kinematical variables and statistics in the experimental measurements,  the most recent phenomenological extractions of Sivers PDFs \cite{Anselmino:2008sga,Anselmino:2012aa} using these data have to rely on several assumptions on the behavior of these PDFs and their flavor dependence, and the resulting fit suffers from significant statistical uncertainties. The future experiments at JLAB12 and the planned EIC will provide more data at higher statistics and wider kinematical range, but clearly more input from a different kind of  measurement will help to relax the approximations and map precisely both the flavor and momentum dependence of Sivers PDFs.

 In our earlier work~\cite{Kotzinian:2014lsa}, we proposed a new approach for measuring the Sivers function in the two-hadron SIDIS process for a transversely polarized target. We presented the results for the relevant cross section calculation that involve the Sivers PDF, where there were two corresponding nonvanishing SSAs with respect to $\sin(\fR-\fS)$ and $\sin(\fT-\fS)$, respectively. Here \vfT, \vfR and $\fS$ are the  azimuthal angles of the total and the relative transverse momenta of the hadron pair and the transverse spin of the nucleon, respectively. In contrast, the leading twist expression for the cross section of the two-hadron SIDIS presented in Ref.~\cite{Bianconi:1999cd} contains only a $\sin(\fT-\fS)$ modulation term. The $\sin(\fR-\fS)$ modulation term is absent there, as well as in the subleading twist expression of Ref.~\cite{Bacchetta:2003vn} for the same cross section, when integrated over the total transverse momentum of the pair.\footnote{Note that our definition of $\fR$, as detailed in the next section, is different than those in Refs.~\cite{Bianconi:1999cd,Bacchetta:2003vn}, while the definitions of $\fT$ are the same.} A thorough discussion of previous work in dihadron fragmentations is presented in Ref.~\cite{Bacchetta:2006un}, while the most general expression for the fully differential two-hadron cross section is presented in Ref.~\cite{Gliske:2014wba}.
 
The measurements of Sivers SSAs in dihadron SIDIS production will provide new information for extracting Sivers PDFs by simply reanalyzing data from single-hadron SIDIS experiments in a different way. Moreover, the dihadron method offers several advantages over single-hadron SIDIS. Namely, in the experiment where $n$  different types of hadrons are detected, there are only $n$ single-hadron and $n(n+1)/2$ dihadron SSAs, providing a good handle for pinpointing the flavor dependence of Sivers PDFs (for not too small $n$). Additionally, if we consider the average multiplicity in the current fragmentation region of each type of hadron $N_i, 1\leq i\leq n$, the statistical uncertainties of single-hadron SSAs for the $i$th type of hadron is inversely proportional to $\sqrt{N_i}$, while for a pair $ij$ ($i\neq j$) it is inversely proportional to $\sqrt{N_iN_j}$. Thus, in the future experiments with large average multiplicities, the dihadron SSAs would be more precisely determined than the single-hadron ones.
 
 In this paper we will present details of these calculations by assuming simple parton-model inspired functional forms for both the unpolarized and Sivers PDFs, as well as the unpolarized dihadron FF (DiFF). Moreover, we will also present the results for single-hadron SSAs in two-hadron SIDIS samples. Finally, similar to~\cite{Kotzinian:2014lsa}, we will use the \LEPT Monte Carlo (MC) event generator~\cite{Ingelman:1996mq}, that has been modified to include the Sivers effect~\cite{Kotzinian:2005zs,Kotzinian:2005zg}, to explore these phenomena in the kinematical region of the COMPASS experiment in a framework that is  independent of the  assumptions employed in deriving the SSAs. We will present an extended set of numerical predictions for both one- and two-hadron SIDIS asymmetries.

This paper is organized in the following way. In the next section we briefly introduce the kinematics of the two-hadron SIDIS process and derive the cross section including the Sivers effect. In Sec.~\ref{SEC_MLEPTO}, we  briefly discuss the modified LEPTO (\MLEPT) MC generator and present the predictions for Sivers SSAs.  In Sec.~\ref{SEC_CONC} we present our conclusions and outlook. Finally, in Appendices we present the detailed derivations of the cross section expressions using simple parametrizations for the PDFs and the DiFFs.

\vspace{-0.5cm}
\section{Sivers effect in two-hadron production}
\label{SEC_SIV_2H}
%
\begin{figure}[tbh]
\begin{center}
\includegraphics[width=1\columnwidth]{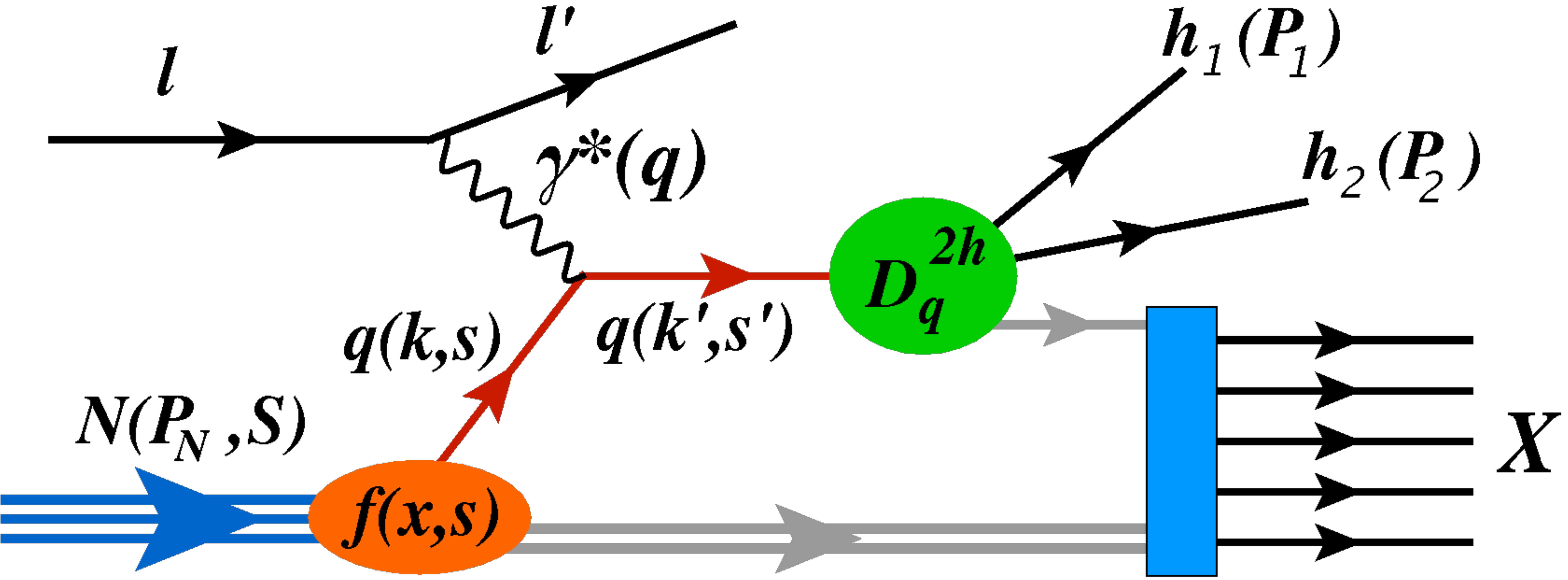}
\caption{The leading order diagram for two-hadron production in the current fragmentation region of SIDIS.}
\label{FIG_DIHADRON_CFR}
\end{center}
\end{figure}
%
\vspace{-0.5cm}
\subsection{Cross section in terms of individual  hadron momenta}
\label{SUB_SEC_XSEC_H1_H2}

 The process of two-hadron electro-production in the current fragmentation region of SIDIS is schematically depicted in Fig.~\ref{FIG_DIHADRON_CFR}.  
 An initial lepton $\ell$ scatters inelastically on a nucleon $N$, where a virtual photon $\gamma^*$ strikes a quark $q$, which then fragments into several hadrons. In the final state, the scattered lepton $\ell'$ and two of the produced hadrons $h_1$ and $h_2$ are detected:
\al{
\label{EQ_2H_SIDIS}
\ell (l ) + N({P_N},S) \to \ell (l') + h_1(P_1) + h_2(P_2) + X\,.
}
 We denote the momenta of the initial and final leptons as $l$ and $l'$ respectively; the momentum of the nucleon $N$ as $P_N$ and the spin $S$; the momenta of the initial and  fragmenting quarks $k$ and $k'$, and the momenta of the detected hadrons as $P_1$ and $P_2$, respectively. In our calculations we use the standard  SIDIS variables:
\al{
\label{EQ_KIN}
&
q=l-l',\,  Q^2 = -q^2, \ x = \frac {Q^2}{2P_N \cdot q},
\\&\non
\, y = \frac {P_N \cdot q}{P_N \cdot l}, \, z_i = \frac{P_N \cdot P_i}{P_N \cdot q}.
}
Here $q$ is the momentum of the virtual photon, $x$ is the usual Bjorken variable for the quark, and $z_i$ is the fraction of the virtual photon's energy in the laboratory system carried by the produced hadron $h_i$.

\begin{figure}[h]
\begin{center}
\includegraphics[width=1\columnwidth]{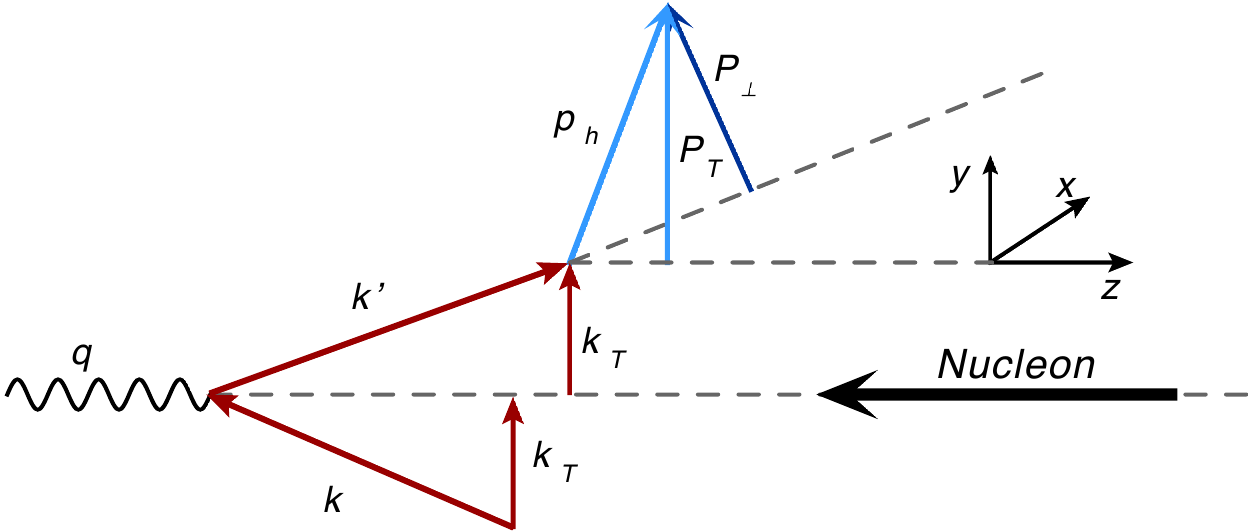}
\caption{$\gamma^*-N$ center-of-mass frame.}
\label{FIG_GAMMA-N_FRAME}
\end{center}
\end{figure}

 Throughout this work we adopt the $\gamma^*-N$ center-of-mass frame, where the $z$ axis is along the direction of the virtual photon momentum $\bfq$. The $x$-$z$ plane is defined by the lepton momenta $\vect{l}$ and $\vect{l}'$. In this frame, we will define the transverse components of the momenta with respect to the $z$ axis with subscript $_T$ and the transverse momenta with respect to the fragmenting quark's direction with subscript $_\perp$, as demonstrated in Fig.~\ref{FIG_GAMMA-N_FRAME}

  Then it is easy to see, using  momentum conservation at the quark-virtual photon vertex, that $\vect{k}'_T=\vect{k}_T$, as $\vect{q}_T\equiv 0$ by definition. Further, we will employ the leading twist approximation for the transverse momenta of the produced hadrons, similar to the single-hadron production case~\cite{Anselmino:2005nn}
\al
{
\label{EQ_PT}
\Pp{1} \approx \PT{1} - z_1 \kT,\\  \non
\Pp{2} \approx \PT{2} - z_2 \kT.
}

In general, the cross section $\sigma^{h_1h_2}$ depends on all the measured independent observables.  In this work we assume that the cross section of the process in Eq.~(\ref{EQ_2H_SIDIS}), where both the hadrons are produced in the current fragmentation region, factorizes into the nonperturbative quark PDF inside of the transversely polarized nucleon, $f_\uparrow^q$; the unpolarized DiFF, $D_q^{h_1h_2}$; and the hard lepton-quark scattering
\al
{
\label{EQ_2H_XSEC}
& \frac{d\sigma^{h_1 h_2}}{dx\, d{Q^2}\, d{\fS}\, dz_1\, dz_2\, d^2\PT{1}\, d^2\PT{2}}  =C(x,Q^2)
\\ \non &
\times \sum_q e_q^2 \int d^2 \kT\  f_\uparrow^q(x, \vect{k}_T)\ D_q^{h_1h_2}(z_1, z_2, \vect{P}_{1\perp}, \vect{P}_{2\perp}),
}
where $C(x,Q^2) =  \frac{\alpha^2}{Q^4}(1+(1-y)^2)$ and $\alpha$ is the fine-structure coupling constant. We also note, that the unpolarized DiFF depends on the energy fractions and transverse components of the produced hadrons with respect to the fragmenting quark $D_q^{h_1h_2}(z_1, z_2, \vect{P}_{1\perp}, \vect{P}_{2\perp})$, as well as $Q^2$.
 
The polarized PDF  $f_\uparrow^q$ depends on both the Bjorken $x$ and the transverse component of the fragmenting quark's momentum $\vect{k}_T$ (and $Q^2$).  It is composed of contributions from the unpolarized $f_1^q(x,k_T)$ and the Sivers PDFs, where the latter arises because of the correlation between the active quark transverse momentum, $\vect{k}_T$, and the transverse polarization of the nucleon, $\mathbf{S}_T$. Then a scalar quantity that describes this correlation can be composed as $\hat{\vect{P}}_N \cdot [\bfk_T \times \bfS]=[\bfS \times \bfk_T]_3=S_T k_T \sin\left(\phi_k-\fS\right)$, with $\hat{\vect{P}}_N$ standing for the unit vector in the direction of $\vect{P}_N$ and subscript 3 denoting the $z$ component of the vector. We can write the polarized PDF explicitly as
\al{
f_\uparrow^q(x,\vect{k}_T)=f_1^q(x,k_T)+\frac{[\bfS \times \bfk_T]_3}{M}f_{1T}^{\perp q}(x,k_T),
}
where $f_{1T}^{\perp q}(x,k_T)$ denotes the Sivers PDF and $M$ is the nucleon mass. Then the cross section of Eq~(\ref{EQ_2H_XSEC}) can be separated into the usual unpolarized part, $\sigma_U$, and a spin-dependent part induced by the Sivers effect, $\sigma_S$:
\al{
\non
& \frac{d\sigma^{h_1h_2}}{dx\, d{Q^2}\, d{\fS}\, dz_1\, dz_2\, d^2\PT{1}\, d^2\PT{2}} = C(x,Q^2) \left({\sigma_U} + {\sigma_{S}}\right),
\\ \non
& \sigma_U = \sum_q e_q^2 \int d^2 \kT \ f_1^q\ D_q^{h_1h_2},
\\
\label{EQ_2H_SIG_SIV}
&\sigma_{S} = \sum_q e_q^2 {\int {{d^2}{\kT}} 
\frac{[\bfS \times \bfk_T]_3}{M} f_{1T}^{ \perp q}\ D_{1q}^{h_1 h_2}}.
}
  It is easy to see using rotational and parity invariance that the most general dependence of $\sigma_S$ on the azimuthal angles $\vf_{1,2}$ (of the transverse momenta $\vect{P}_{1,2T}$) and~$\fS$ is given by two "Sivers-like" terms
\al{
\label{EQ_2H_SIV_GEN}
\sigma_{S} 
= S_T\left(\sigma_1\frac{P_{1T}}{M}  \sin(\vf_1-\fS) \right.&
\\ \non 
+
&\left. \sigma_2\frac{P_{2T}}{M} \sin(\vf_2-\fS)\right),
}
where $\sigma_{S}, \sigma_1$ and $\sigma_2$ depend on $x,Q^2,z_1,z_2,P_{1T},P_{2T}$ and $\vect{P}_{1T}\cdot\vect{P}_{2T}$ [or $\cos(\vf_1-\vf_2)$]. 

Here we use parton model parametrizations of the relevant PDFs and DiFFs to derive explicit expressions for $\sigma_1$ and $\sigma_2$. Thus we adopt the usual factorization between the intrinsic transverse momentum and the energy fraction dependences for these functions. Further, we assume for the PDFs a Gaussian dependence on the transverse momentum
\al{
f_{1}^q(x,k_T) = f_{1}^{q}(x) \, \frac{1}{\pi \mu_0^2} \, e^{-{\kt^2}/{\mu_0^2}},
\label{EQ_UNPOL_PDF}
}
where $\mu_0^2=\langle \kt^2 \rangle$ is the mean value of the intrinsic transverse momentum of a quark inside the nucleon. In practice $\mu_0^2$ is a function of the quark flavor, $x$ and $Q^2$ (see, e.g.. Ref.~\cite{Matevosyan:2011vj}). We adopt a similar parametrization for the Sivers PDF from Ref.~\cite{Anselmino:2005nn}:
\al{
f_{1T}^{ \perp q}(x,k_T) = f_{1T}^{ \perp q}(x) \, \frac{1}{\pi \mu_{S}^2}
e^{-{\kt^2}/\mu_{S}^2} \,.
\label{EQ_SIV_PDF}
}

The unpolarized DiFF $D_{1q}^{h_1h_2}$ should depend only on $z_1,z_2,P_{1\perp},P_{2\perp}$ and $\Pp{1}\cdot \Pp{2}$ (again because of rotational invariance). Then we can adopt the following form for it:
\al{
\label{EQ_UNPOL_DIFF}
&D_{1q}^{h_1h_2}\left(z_1,z_2,P_{1\perp},P_{2\perp},\vect{P}_{1\perp}\cdot\vect{P}_{2\perp}\right)=
\\&\non
D_{1q}^{h_1h_2}(z_1,z_2)\frac{1}{\pi^2 \nu_1^2 \nu_2^2} \, e^{-P_{1\perp}^2/\nu_1^2-P_{2\perp}^2/\nu_2^2}
\left(1+c \Pp{1} \cdot \Pp{2}\right),
}
where the term $c{(\Pp{1} \cdot \Pp{2})}$ takes into account the transverse momentum correlations experimentally established in fragmentation (see, e.g. Ref.~\cite{Arneodo:1986yc}). Note that in general the parameters $\nu_1^2$, $\nu_2^2$ and $c$ could depend on $Q^2,z_1$ and $z_2$, the flavor of the fragmenting quark $q$, and the type of the produced hadrons, similar to the case of the single-hadron fragmentation~\cite{Matevosyan:2011vj}. Here we omit the corresponding indices for such dependencies for brevity.

 Using the parametrizations for the PDFs and DiFFs of Eqs.~(\ref{EQ_UNPOL_PDF}-\ref{EQ_UNPOL_DIFF}), we can  easily perform the integral over $\vect{k}_T$ in Eq.~(\ref{EQ_2H_SIG_SIV}), with resulting expressions for the unpolarized and spin-dependent cross section terms:
\al{
\label{EQ_2H_XSEC_12RES}
&\sigma_U = \sum\limits_q e_q^2 f_1^q(x)D_{1q}^{h_1h_2}(z_1,z_2)C_0^{h_1h_2},
\\&\non
\sigma_1 = \sum\limits_q e_q^2 f_{1T}^{\perp q}(x)D_{1q}^{h_1h_2}(z_1,z_2)C_1^{h_1h_2},
\\&\non
\sigma_2= \sum\limits_q e_q^2 f_{1T}^{\perp q}(x)D_{1q}^{h_1h_2}(z_1,z_2)C_2^{h_1h_2},
}
where the expressions for the factors $C_{i}^{h_1h_2}$ are presented in Appendix~\ref{SEC_APP_P12}.

 The results of the experimental measurements of the two-hadron SIDIS production cross section are often presented after integration over the azimuthal angle of one of the transverse momenta and some range of its magnitude. In Appendix~\ref{SEC_APP_P12_1H} we calculate these integrals of the two-hadron cross section over $\vf_1$ or $\vf_2$ 
\al{
&\label{EQ_2H_XSEC_INT1}
\frac{d \sigma^{h_1h_2 } } {P_{1T}\ d P_{1T}\ d^2 \PT{2} } = C(x,Q^2)
\\\non&
\times \left[ \sigma_{U,0} +  S_T \left( \frac{P_{1T}}{2M} \sigma_{1,1}+\sigma_{2,0} \frac{P_{2T}}{M} \right) \sin(\vf_2-\vf_S)  \right],
\\&\label{EQ_2H_XSEC_INT2}
\frac{d \sigma^{ h_1h_2 } }
{d^2 \PT{1}\ P_{2T}\ d P_{2T}}
= C(x, Q^2) 
\\\non&
\times 
\left[ \sigma_{U,0} +    S_T \left( \frac{P_{1T}}{M} \sigma_{1,0}+\sigma_{2,1} \frac{P_{2T}}{2M} \right) \sin(\vf_1-\vf_S)  \right],
}
where $\sigma_{U,0}$ and $\sigma_{1(2),0(1)}$ denote the moments with respect to $\cos(n(\vf_1-\vf_2))$ Fourier expansion of the corresponding cross section terms of \Eq{EQ_2H_XSEC_12RES}. Examining the explicit expressions for these moments presented  in Appendix~\ref{SEC_APP_P12_1H}, it is easy to conclude that, in general, Sivers modulations of one hadron do not vanish when we integrate over the azimuthal angle of the other hadron.

\subsection{Cross section in terms of relative and total hadron momenta}

 It is often convenient to use linear combinations of $\PT{1}$ and $\PT{2}$ as independent transverse momentum variables. In literature the total transverse momentum, $\vT$, and the half  of the relative transverse momentum of the hadron pair, $\vect{R}$ are  conventionally chosen~\cite{Bianconi:1999cd, Airapetian:2008sk,Adolph:2012nw}
\al
{
\label{EQ_T_R}
&
\vT=\vect{P}_{1T}+\vect{P}_{2T},
\\&
\vect{R}=\frac{1}{2}\left(\vect{P}_{1T}-\vect{P}_{2T}\right),
}
where the corresponding azimuthal angles are denoted as $\vf_T$ and $\vf_R$. We can easily transform the the expressions for the cross section in Eqs.~(\ref{EQ_2H_SIG_SIV},\ref{EQ_2H_SIV_GEN}) by noting that the Jacobian of the transformation $\partial(\PT{1}, \PT{2})/\partial(\vT,\vect{R})$ is simply $1$.  Then we can express the cross section as
\al{
\label{EQ_2H_X_SEC_RT}
&\frac{d\sigma^{h_1 h_2}}{d^2\vT\, d^2\vect{R}\,} = C(x,Q^2) 
\left[\sigma_U \vphantom{\frac{1}{1}} \right.
\\ \non
&
\left.+ 
S_T \left(\sigma_T\frac{P_T}{M}\sin(\vf_T-\fS) + \sigma_R\frac{R}{M}\sin(\vf_R-\fS)\right)\right],
}
where $\sigma_T = \frac{1}{2}\left(\sigma_1+\sigma_2\right)$,  $\sigma_R = \sigma_1-\sigma_2$, and we suppressed  the dependence of the cross section on the rest of the variables for brevity. The details of the derivation and the explicit expressions for $\sigma_T$ and $\sigma_R$ can be found in Appendix~\ref{SEC_APP_RT}.  We note that the structure functions $\sigma_U,\ \sigma_T$ and $\sigma_R$ depend on $x,\ Q^2, \ z_1, \ z_2, \ P_T, \ R$ and  $\vT\cdot\vect{R}=P_T R\cos(\vf_T-\vf_R)$. Moreover, it is easy to convince oneself by examining the expression in Eq.~(\ref{EQ_APP_C_SIGMA_R}), that in general $\sigma_R \neq 0$. This can be ensured, for example, by choosing asymmetric cuts on the minimum values of $z_1$ and $z_2$. 

 Keeping in mind the experimental extractions of the Sivers asymmetries, it is useful to calculate the cross section in Eq.~(\ref{EQ_2H_X_SEC_RT}) after integrating over  the azimuthal angle of the relative or total transverse momentum, $\fR$ or $\fT$ respectively:
\al{
\label{EQ_2H_X_SEC_INT_R}
\frac{d\sigma^{h_1h_2}}{d^2\vT RdR} &= C(x, Q^2)
\left[
\sigma_{U,0}  \vphantom{\frac{1}{1}} \right.
\\ \non
&\left. +
S_T \left(\frac{P_T}{M}\sigma_{T,0}+\frac{R}{2M}\sigma_{R,1} \right)\sin(\vf_T-\fS)
\right],
\label{EQ_2H_X_SEC_INT_T}
\\ 
\frac{d\sigma^{h_1h_2} }{P_T dP_T d^2\vect{R}} &= C(x, Q^2)
\left[
 \sigma_{U,0} 
\vphantom{\frac{1}{1}} \right.
\\ \non&
\left. +
S_T \left(\frac{P_T}{2M}\sigma_{T,1}+\frac{R}{M}\sigma_{R,0} \right)\sin(\vf_R-\fS)
\right],
}
where $\sigma_{U,i}, \sigma_{T,i}$ and $\sigma_{R,i}$ are the zeroth ($i=0$) and the first ($i=1$) harmonics of the $\cos(n(\vf_T-\vf_R))$ Fourier expansions of the corresponding structure functions. The explicit expressions for these functions are presented in Appendix~\ref{SEC_APP_RT_1H}. It can be seen clearly from the expressions in Eqs.~(\ref{EQ_APP_D_SIGMA_U0},\ref{EQ_APP_D_SIGMA_T0},\ref{EQ_APP_D_SIGMA_T1},\ref{EQ_APP_D_SIGMA_R0},\ref{EQ_APP_D_SIGMA_R1}), that in general the Sivers effect is nonzero in Eqs.~(\ref{EQ_2H_X_SEC_INT_R},\ref{EQ_2H_X_SEC_INT_T}). Moreover, both the $\sin(\fT-\fS)$ and $\sin(\fR-\fS)$ modulations have contributions from both the $\sigma_{T}$ and $\sigma_{R}$ unintegrated  cross section terms.

\subsection{Sivers effect for $\vect{R}$ with Pavia convention}

 It has been noted in the Introduction, that in previous work of Refs.~\cite{Bianconi:1999cd,Bacchetta:2003vn}, modulations with respect to the azimuthal angle of the relative transverse momenta are absent in the dihadron cross section.  This seeming contradiction with our results of \Eq{EQ_2H_X_SEC_INT_T} can be resolved by examining the differences in the definitions of the relative transverse momenta. In Pavia notation of Refs.~\cite{Bianconi:1999cd,Bacchetta:2003vn}, the relative transverse momentum of the hadron pair, $\vect{R}^{P}$, is defined as the component of $\vect{R}$ transverse to $\vect{P}_h$:
\al
{
\vect{R}^{P} \equiv \vect{R} - (\vect{R} \cdot \hat{\vect{P}}_h) \hat{\vect{P}}_h = \vect{R} - \frac{(\vect{R} \cdot \vect{P}_h) \vect{P}_h}{P_h^2}
}

In the leading order approximation in the deep inelastic scattering (DIS) regime, where we neglect the small transverse momenta and mass terms compared to $Q$, we can make the approximation 
\al
{
\vect{P}_i^2 = E_i^2 -m_i^2 \approx z_{i}^2 \vect{k}'^2,
}

Then
\al
{
\vect{R}^P& = \vect{R} -\frac{z_{1}^2 - z_{2}^2}{(z_{1}+z_{2})^2} \frac{\vect{P}_h}{2} 
= \xi_2 \vect{P}_1- \xi_1 \vect{P}_2,
\\
\xi_i &\equiv \frac{z_i}{z_1+z_2}.
}

Finally
\al
{
\vect{R}^P_T &=
 \xi_2 \vect{P}_{1T} -  \xi_1 \vect{P}_{2T}
 =\xi_2 \vect{P}_{1\perp} - \xi_1 \vect{P}_{2\perp} .
 }

Thus $\vect{R}^P_T$, at the leading order approximation, is completely disconnected form the fragmenting quark's transverse momentum $\kT$ , thus it should produce no Sivers asymmetry in the corresponding cross section. Further details and  model-independent expressions for the cross sections of the dihadron production as functions of either choices of $\vect{R}$ are presented in Ref.~\cite{Kotzinian:2014hoa}.

\section{Modified \LEPT (\MLEPT) including the Sivers effect at leading order}
\label{SEC_MLEPTO}

\begin{figure}[tb]
\centering 
\subfigure[] {
\includegraphics[width=\ImS]{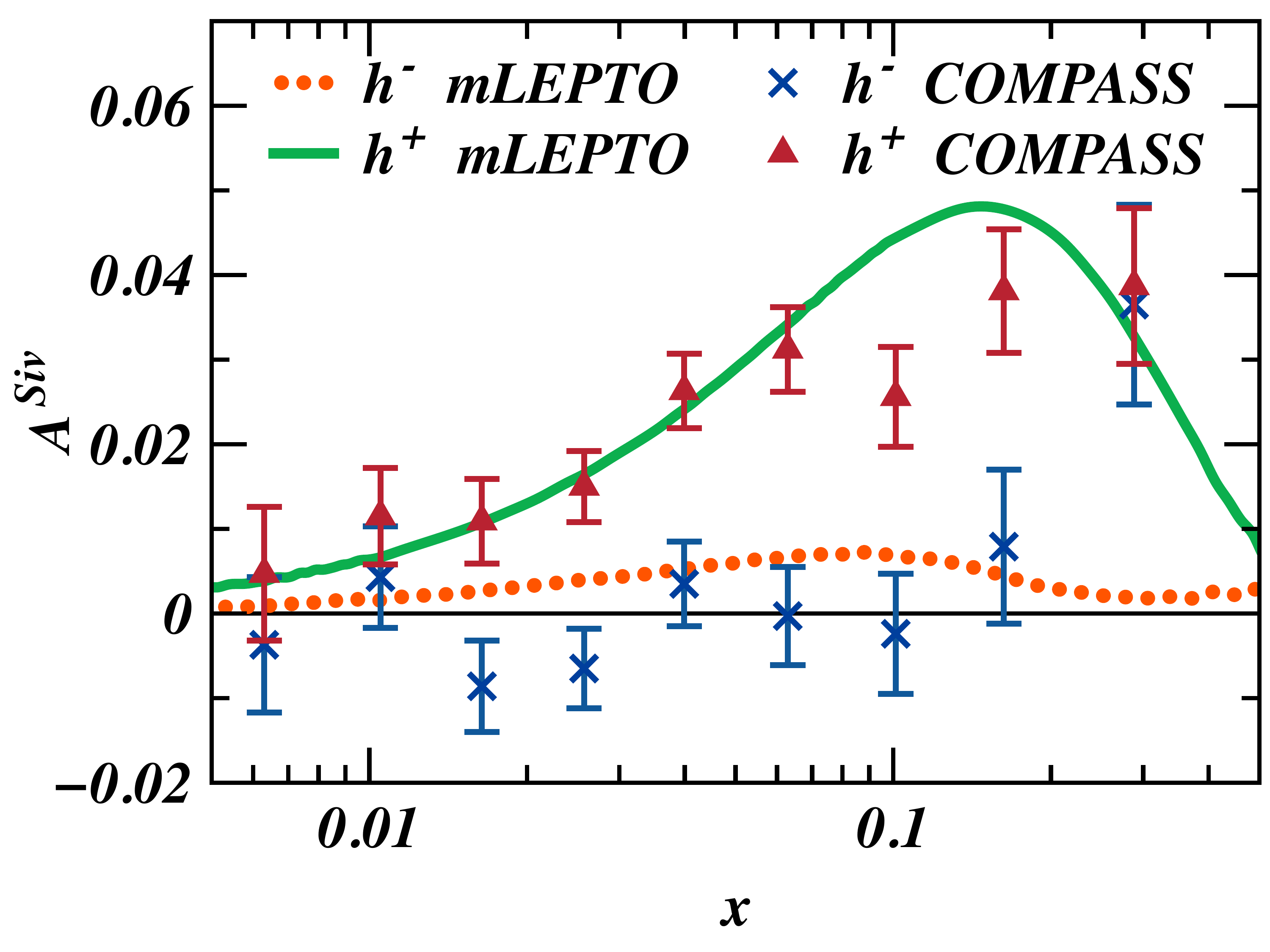}}
\\\vspace{-0.2cm}
\subfigure[] {
\includegraphics[width=\ImS]{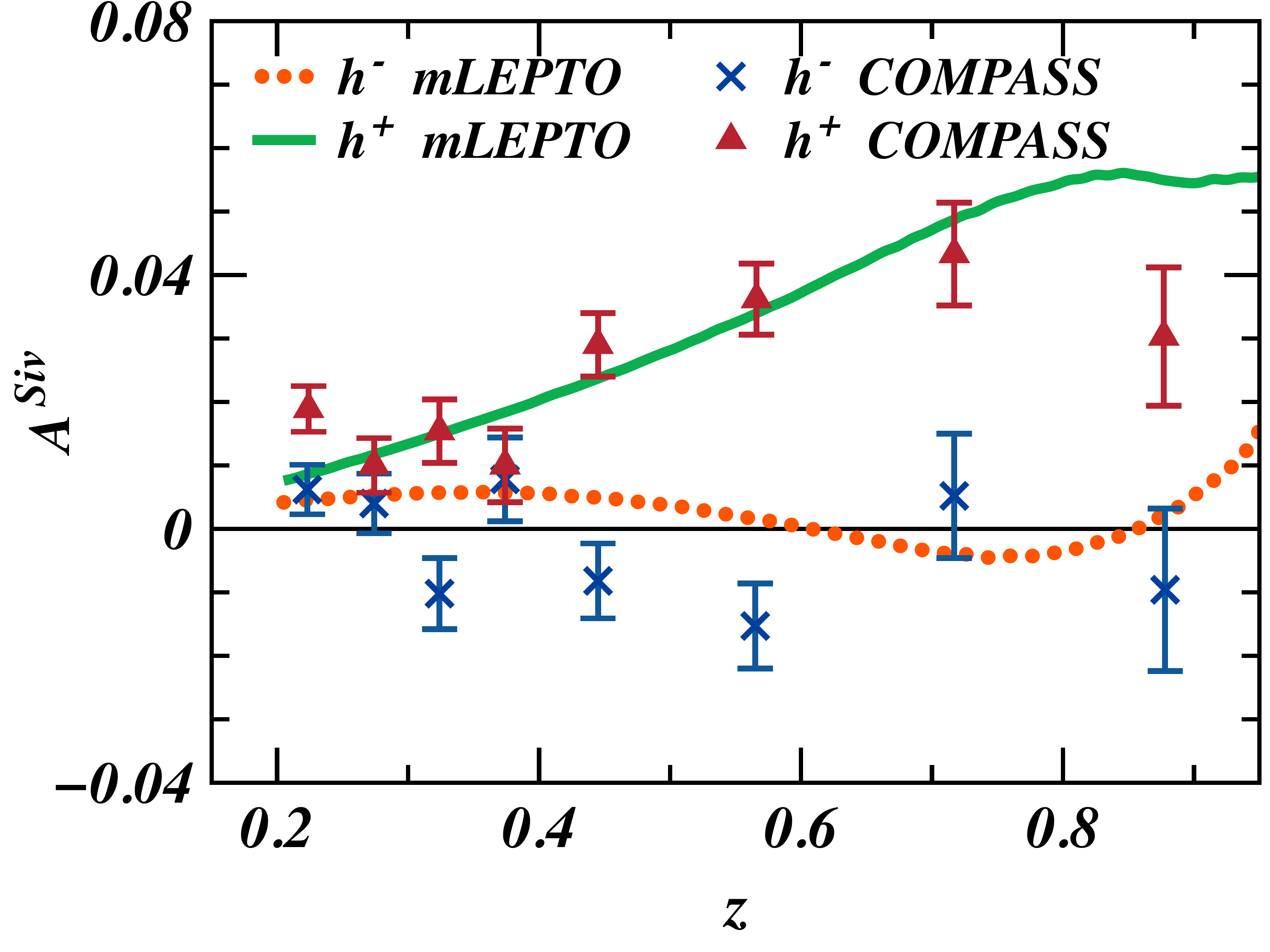}
}
\\\vspace{-0.2cm}
\subfigure[] {
\includegraphics[width=\ImS]{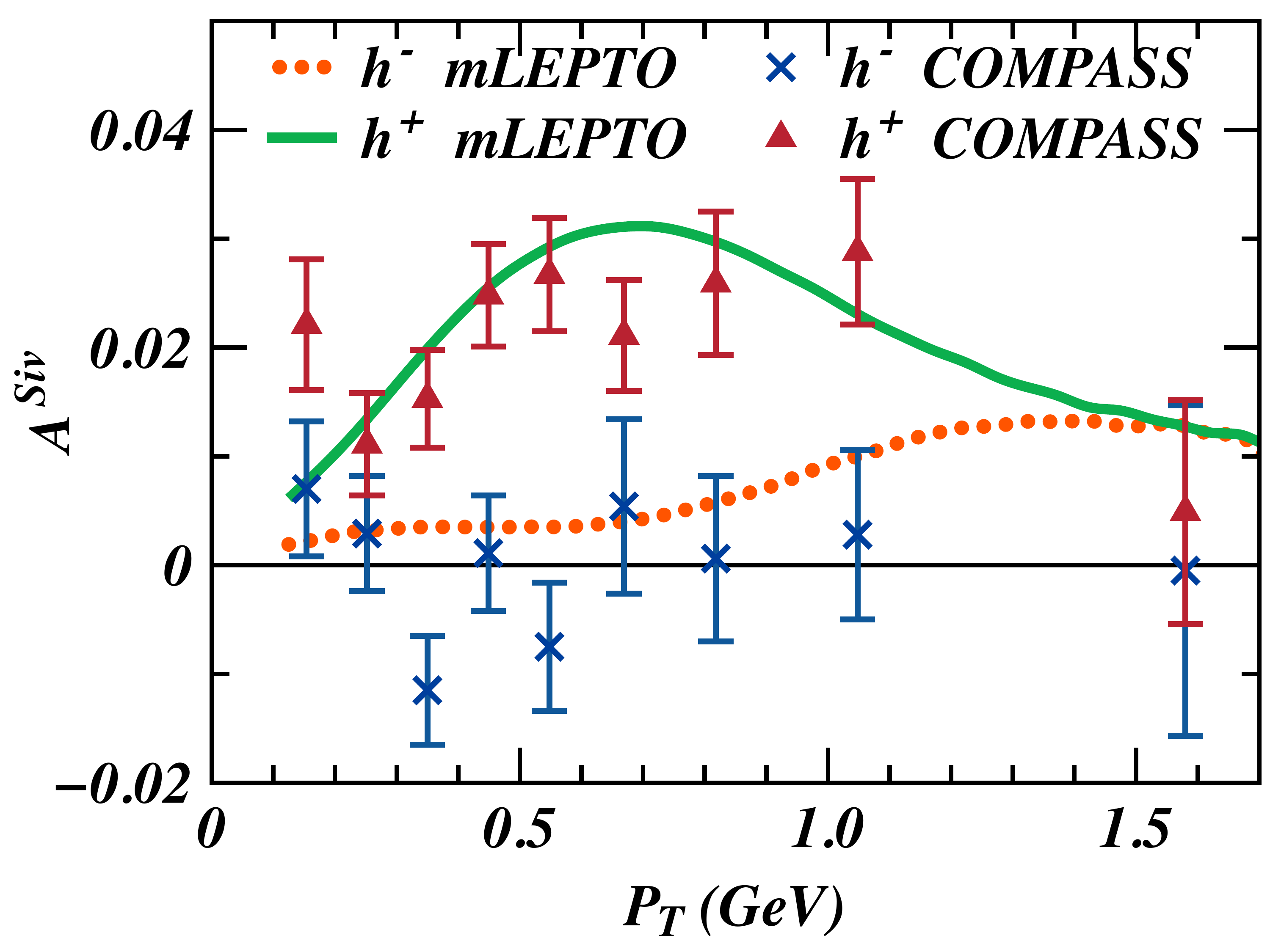}
}
\\\vspace{-0.2cm}
\caption{COMPASS results for Sivers asymmetry in a charged (triangles for positive and crosses for negative) hadron production off proton target, compared to those from \MLEPT (bands), for  $x$ (a), $z$ (b), and  $P_T$ (c) dependencies. The width of each band indicates the statistical accuracy of our simulations and does not include the uncertainties of the PDFs.}
\label{PLOT_SIV_1H}
\end{figure}

 Monte Carlo event generators are an important tool in both theoretical an experimental studies of high energy DIS reactions. At present, however, none of the these event generators implement the simulations of the experimentally observed azimuthal asymmetries for hadron production in SIDIS on either unpolarized or polarized targets. Here our goal is to use an event generator to test the existence of the azimuthal modulations induced by the Sivers effect in two-hadron SIDIS,  to support the parton model derivations for the relevant cross section performed in the previous section. Thus, we chose one of the widely used unpolarized event generators,  \LEPT~\cite{Ingelman:1996mq}, and modified the code (\MLEPT) to include both Cahn and Sivers azimuthal modulations of the transverse momentum of the active quark before hard scattering and hadronization~\cite{Kotzinian:2005zs,Kotzinian:2005zg}. Here we first employ \MLEPT to calculate the SSAs for the Sivers effect in the single-hadron SIDIS process in the kinematical region of the COMPASS experiment, using the empirical parametrizations of Eq.~(\ref{EQ_SIV_PDF}) of the Sivers PDFs from Ref.~\cite{Anselmino:2012aa} that were fitted to the experimental data of Refs.~\cite{Airapetian:2009ae,Adolph:2012sp}. This allows us to validate \MLEPT by comparing the dependence of these SSAs for both positively and negatively charged hadrons to those extracted in experiment and slightly adjust the parameters (within their uncertainties) of the Sivers PDFs to achieve the best match. Further, we use \MLEPT to calculate the SSAs in two-hadron SIDIS induced by the Sivers effect for the kinematics of the COMPASS experiment. We note, that the hadronization in the Lund model, implemented in \LEPT, is  different from the independent FFs of the parton model factorization theorem. Nevertheless, these two approaches produce similar results for the current fragmentation region of COMPASS kinematics~\cite{Kotzinian:2004xq}.
 
In the mLPETO simulations used in this article we generated $10^{11}$ DIS events in the kinematical region of the the COMPASS experiment~\cite{Adolph:2012sp}, where $E_\mu=160~\Ge$ muons were scattered off the $NH_3$ target. The following kinematic cuts are applied:  $Q^2>1~\Gs$, $0.1<y<0.9$, $0.03<x<0.7$, $W>5~\Ge$.  The single-hadron SSAs calculated in this simulations then can be directly compared with those measured in Ref.~\cite{Adolph:2012sp}. 

 First we examine the SSAs for the single-hadron SIDIS production off transversely polarized proton target, induced by the Sivers effect. 
 We present the results of our \MLEPT simulations with only a single $\sin(\varphi-\varphi_S)$  modulation present, where $\varphi=\varphi_h$ for one hadron production, and $\varphi=\fR$ or $\varphi=\fT$ for two-hadron production (see Eqs.~(\ref{EQ_2H_X_SEC_INT_R},\ref{EQ_2H_X_SEC_INT_T})). Then the cross section can be written as
\al{
d\sigma \propto \sigma_U+S_T \tilde{\sigma}_S \sin(\varphi-\varphi_S),
}
where $\sigma_U, \tilde{\sigma}_S$ are the corresponding unpolarized and Sivers cross section terms, and the asymmetry is defined as 
\al{
A_{Siv}=\frac{\tilde{\sigma}_S}{\sigma_U}.
}

 The \MLEPT results for both positively and negatively charged hadron are depicted on the plots in Fig.~\ref{PLOT_SIV_1H} as functions of:  (a) the Bjorken $x$, (b) the produced hadrons' energy fraction $z$, and (c) the transverse momentum $P_T$. The COMPASS results of Ref.~\cite{Adolph:2012sp} are also depicted here, and \MLEPT describes them well. Here we imposed additional kinematic cuts $P_T>0.1~\Ge$ and $z>0.2$, following the procedure used by the COMPASS collaboration in Ref.~\cite{Adolph:2012sp}. 
 
\begin{figure}[tb]
\centering 
\includegraphics[width=\ImS]{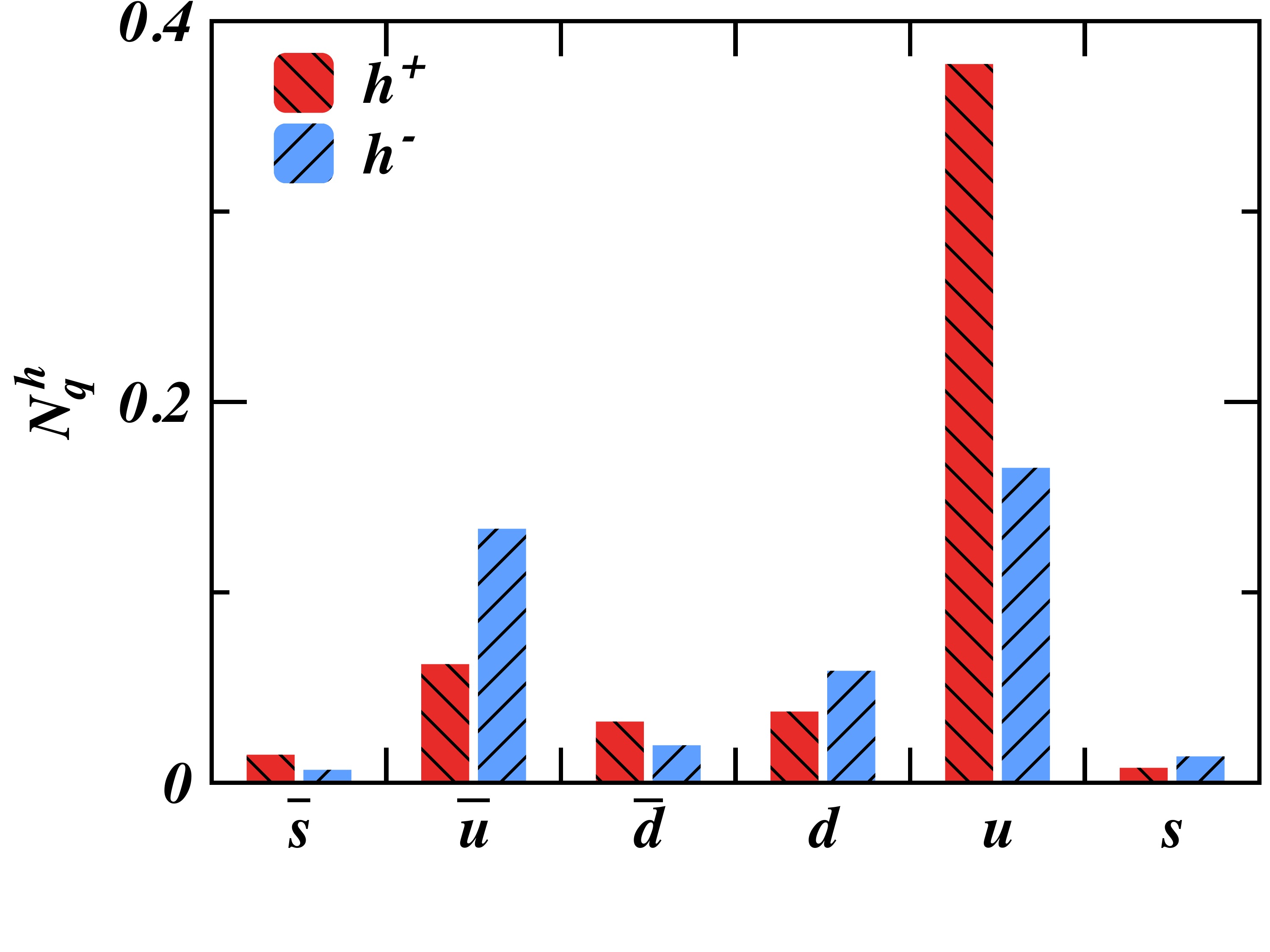}
%
\\\vspace{-0.2cm}
\caption{\MLEPT predictions for the relative rates for the flavor of the struck quark that produces positively (in red) and negatively (in blue) charged hadrons in MC events with all the relevant kinematical cuts.}
\label{PLOT_SIV_QUARK_ID}
\end{figure}

 \MLEPT allows one to determine the flavor of the struck quark in each SIDIS event, which provides very useful information for the phenomenological analysis of the results. The plots in Fig.~\ref{PLOT_SIV_QUARK_ID} depict the results for the relative rates for the struck quark flavor in events that produce a given type hadron satisfying the kinematic cuts imposed in our analysis. Here we find that the positively charged hadrons, $h^+$ in red,  are predominately produced by the $u$ quarks, while the negatively charged hadrons, $h^-$ in blue, are produced almost equally by both $u$ and $\bar{u}$ quarks, as well as significantly by the $d$ quark. These results are easy to interpret by considering the relative magnitudes of the quark PDFs and the fragmentation functions in the kinematics of this analysis.
\begin{figure}[tb]
\centering 
\subfigure[] {
\includegraphics[width=\ImS]{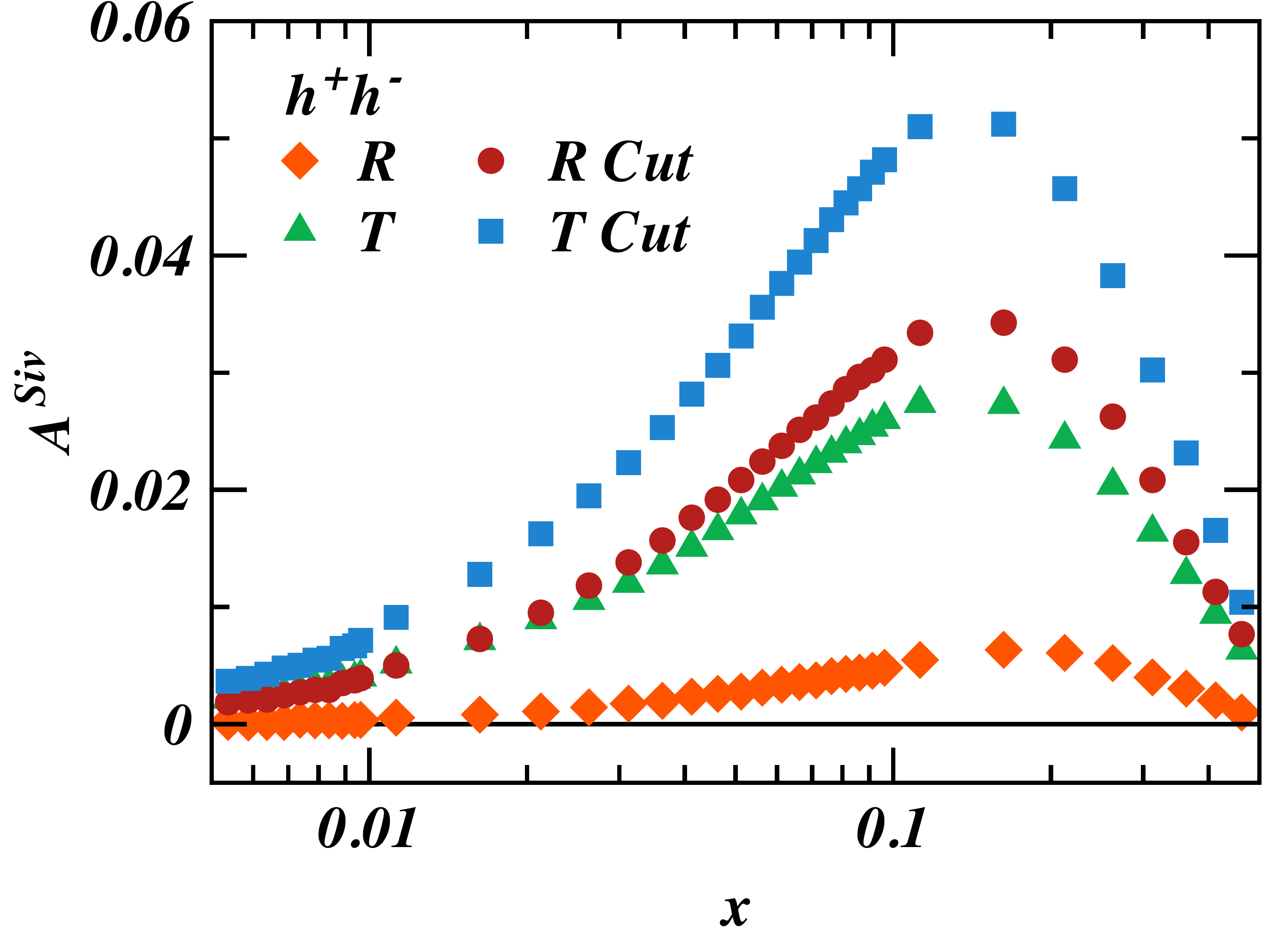}}
\\\vspace{-0.2cm}
\subfigure[] {
\includegraphics[width=\ImS]{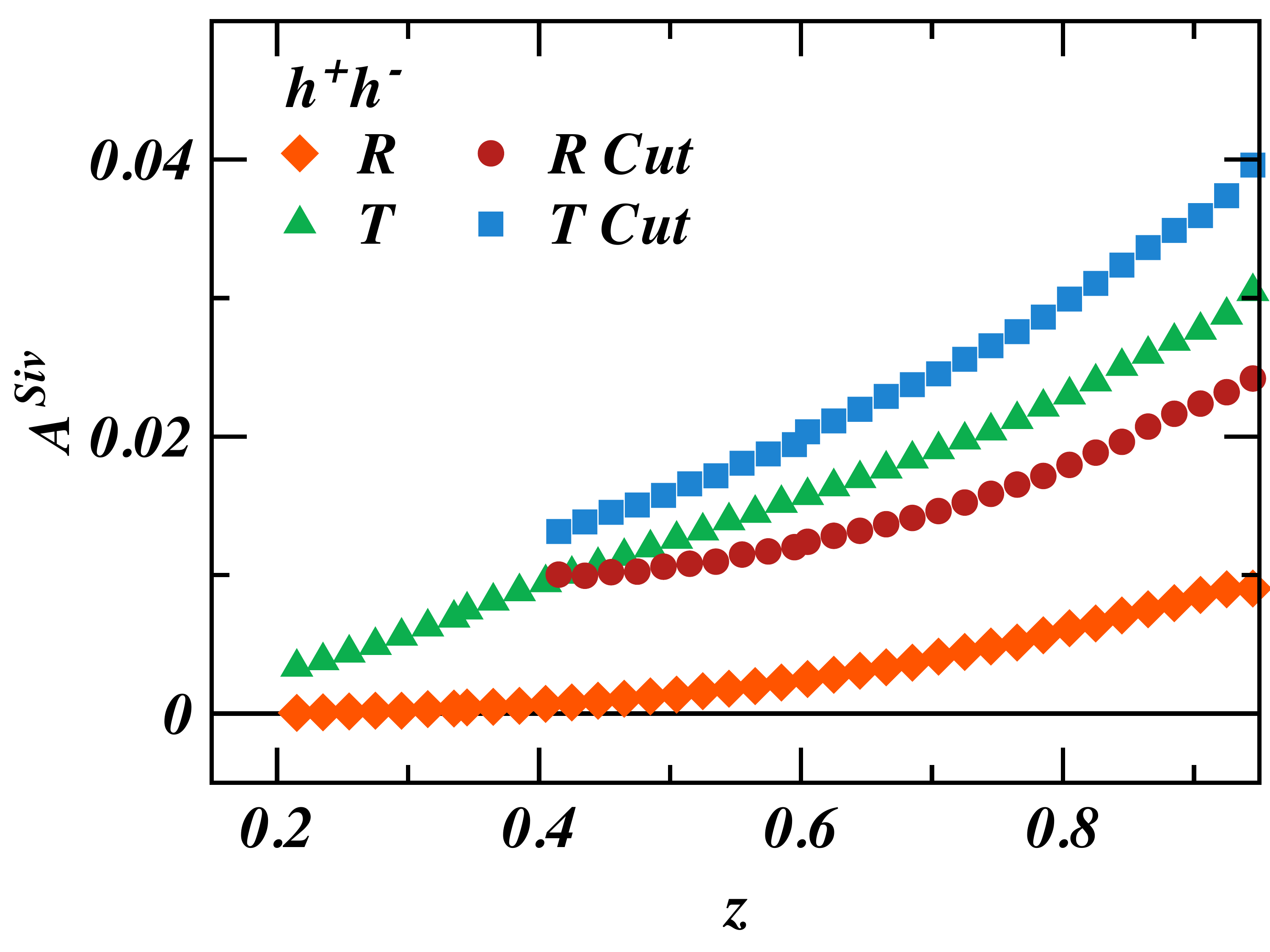}
}
\\\vspace{-0.2cm}
\caption{\MLEPT predictions for the dependence of the Sivers asymmetry on: (a) $x$, and (b) the total energy fraction $z$, in oppositely charged hadron pair production off proton target for both $\fR$ and $\fT$ asymmetries integrated over $\vect{P}_T$ and $\vect{R}$, respectively. The bands labeled "Cut" are the results with the additional cut on the positively charged hadron's momentum, as described in the text.}
\label{PLOT_SIV_2H_MIPL_X_Z}
\end{figure}

\begin{figure}[tb]
\centering 
\subfigure[] {
\includegraphics[width=\ImS]{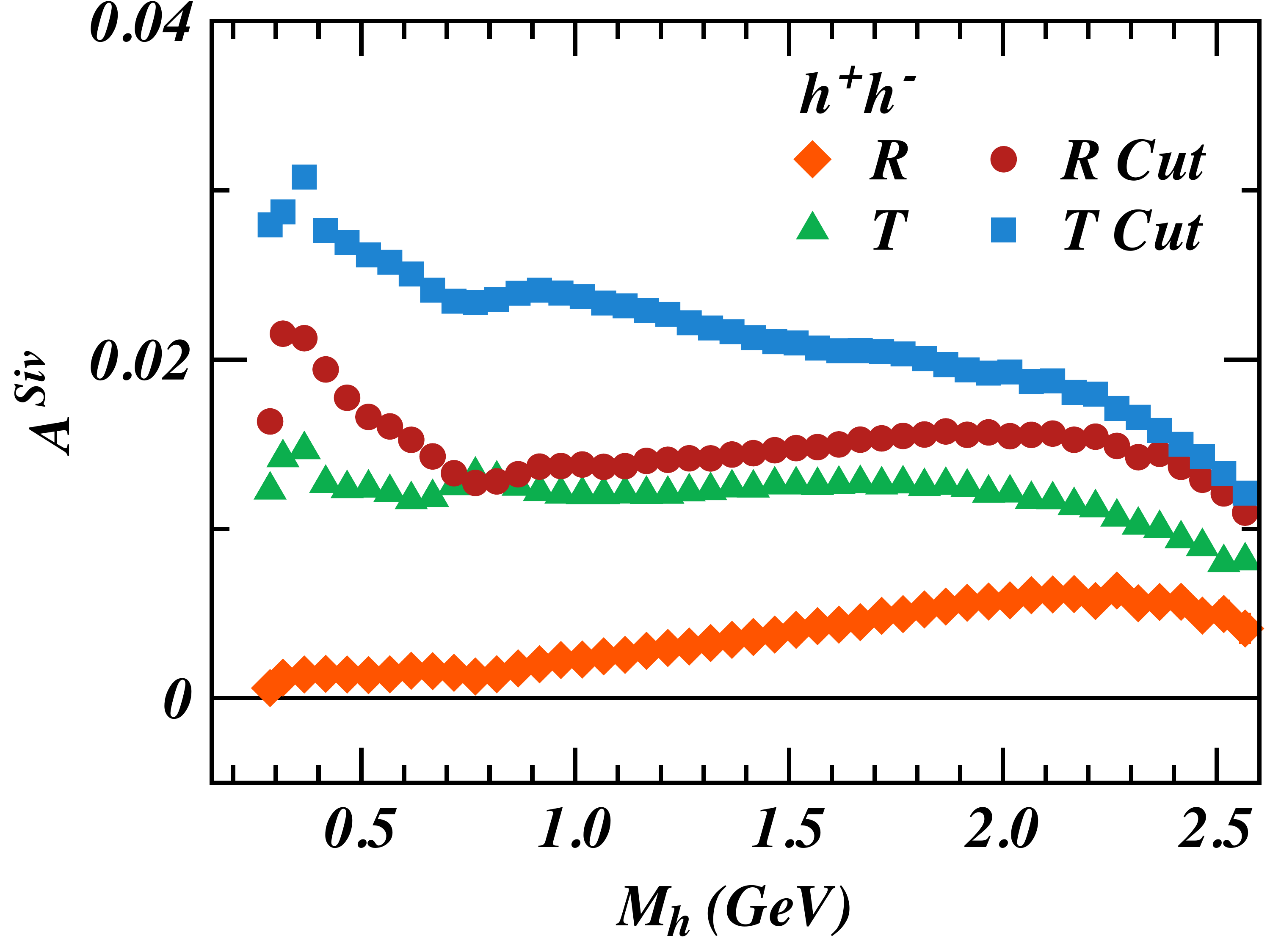}}
\\\vspace{-0.2cm}
\subfigure[] {
\includegraphics[width=\ImS]{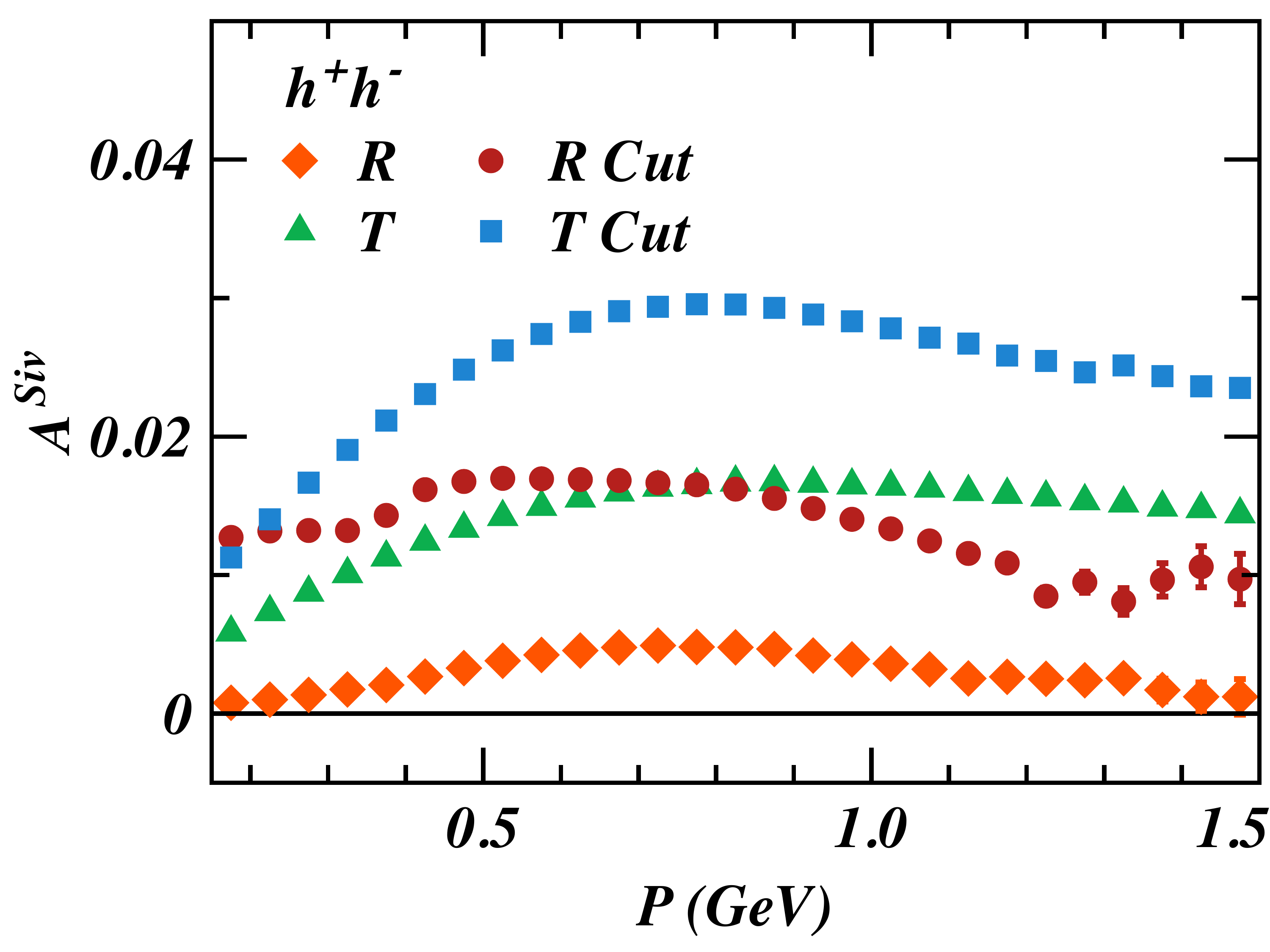}
}
\\\vspace{-0.2cm}
\caption{\MLEPT predictions for the dependence of the Sivers asymmetry on (a) $M_h$ and (b) the magnitude of the relevant momentum $P$, in oppositely charged hadron pair production off proton target for both $\fR$ and $\fT$ asymmetries integrated over $\vect{P}_T$ and $\vect{R}$, respectively. The bands labeled "Cut" are the results with the additional cut on the positively charged hadron's momentum, as described in the text.}
\label{PLOT_SIV_2H_MIPL_MH_PT}
\end{figure}

 Next, we present on the plots in Fig.~\ref{PLOT_SIV_2H_MIPL_X_Z} our results  predicting the dependence of the Sivers SSAs for a SIDIS production of a hadron pair with opposite charges $h^+h^-$. In the analysis of the MC generated events, the hadron labelled "1" (the first hadron) is chosen as $h^+$ and the hadron labelled "2" ( the second hadron) as $h^-$ and we extract the SSAs corresponding to both $\sin(\fT-\vf_S)$ and $\sin(\fR-\vf_S)$ modulations (labelled "T" and "R" on the plots). The dependence of these SSAs on the quark $x$ are depicted in subfigure (a)  and on the total energy fraction of the pair $z=z_1+z_2$ in subfigure (b).  Further, the plots in Fig~\ref{PLOT_SIV_2H_MIPL_MH_PT} depict the results for the SSAs as functions of the invariant mass $M_h$ in subfigure (a), and the transverse momentum corresponding to the type of the modulation in subfigure (b).  Again, we imposed the following kinematic cuts  on the momenta of the hadrons $P_{1(2)T}>0.1~\Ge$, $z_{1(2)}>0.1$. The SSAs, especially corresponding to the R modulations are suppressed in comparison with  the single-hadron case, and thus might be harder to access experimentally. We have checked that in our MC sample the distributions of the first and the second hadrons are practically identical as functions of the transverse momentum and fractional energy. This observation can explain the smallness of the R-type Sivers asymmetry as a function of $x$, since if $\nu_1=\nu_2$ and $\langle z_1\rangle=\langle z_2\rangle$,  then $ \beta_{S1}=\beta_{S2}$ [see Eq.~(\ref{EQ_APP_C_SIGMA_R})]. As expected from the results in the previous section, these SSAs can be enhanced by imposing asymmetric cuts on the momenta of the hadrons in a pair.  The bands labeled "Cut" in Fig.~\ref{PLOT_SIV_2H_MIPL_X_Z} depict the results with cuts on the momentum of the positively charged hadron: $z_1>0.3$ and $P_{T1}>0.3~\Ge$. The results exhibit a significant enhancement for both T- and R-type SSAs, making them comparable in magnitude to those for the single-hadron SIDIS production.
 
\begin{figure}[tb]
\centering 
\subfigure[] {
\includegraphics[width=\ImS]{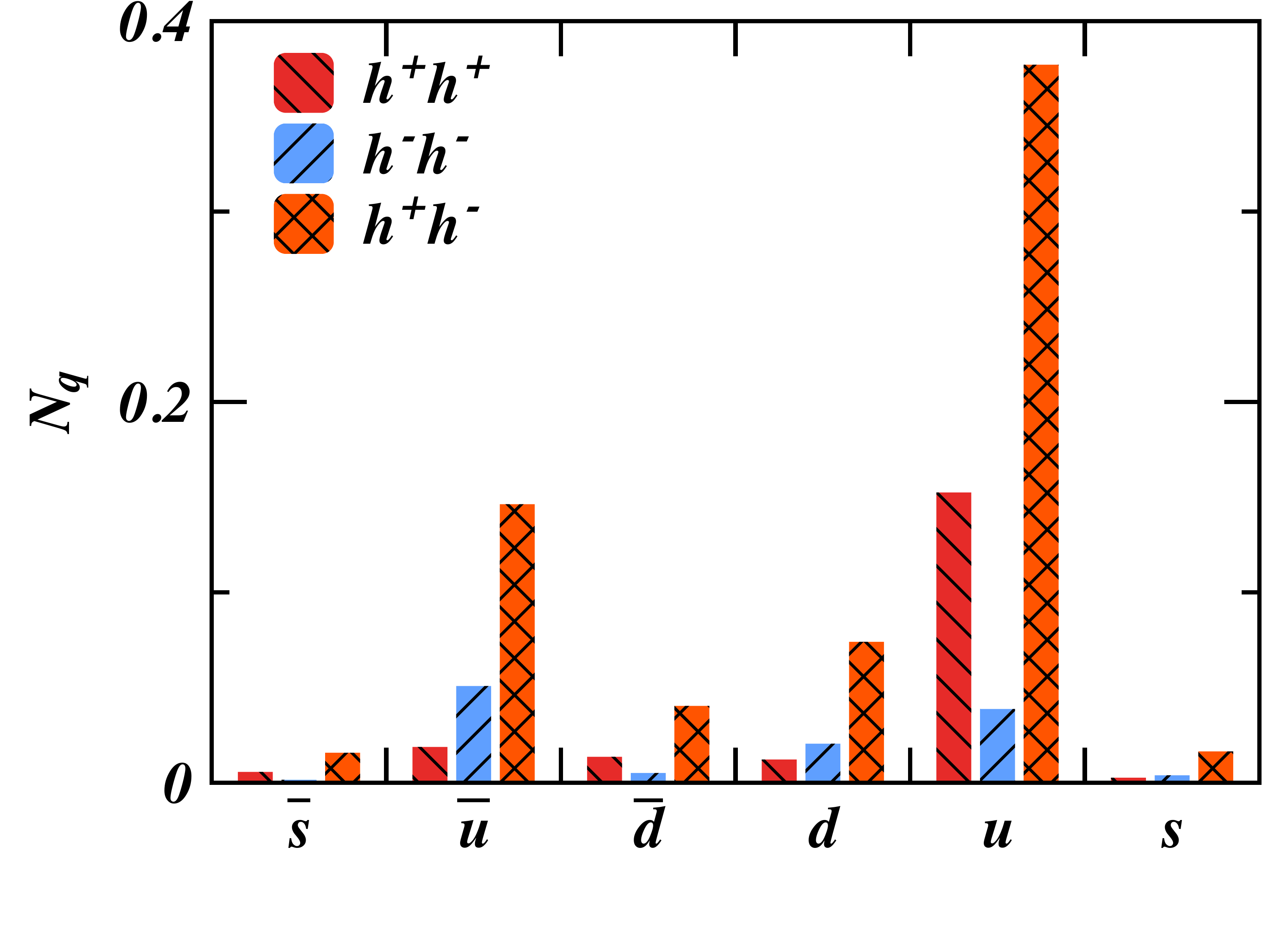}
}
\\\vspace{-0.2cm}
\subfigure[] {
\includegraphics[width=\ImS]{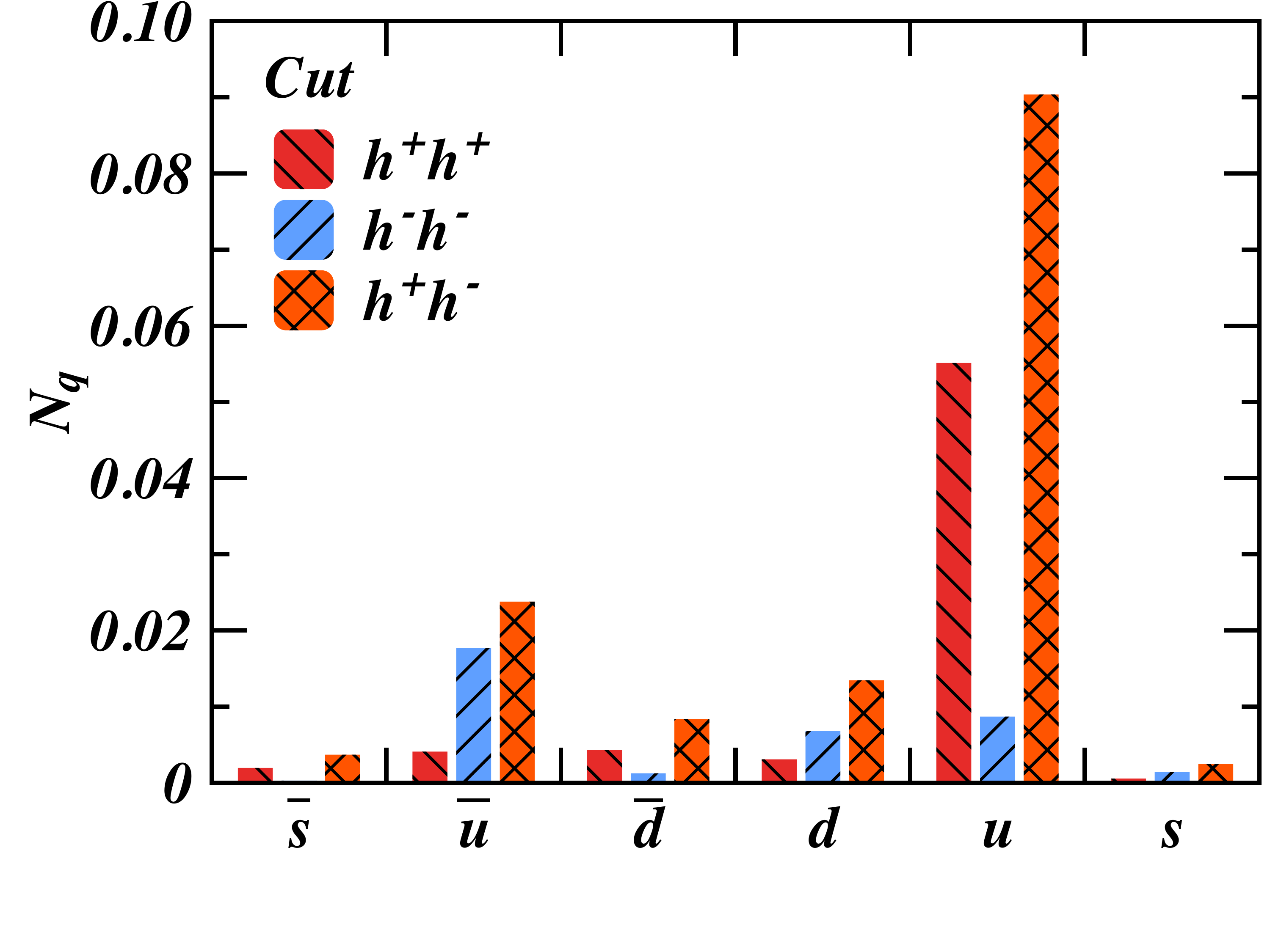}
}
\\\vspace{-0.2cm}
\caption{\MLEPT predictions for the relative rates for the flavor of the struck quark that produces a pair of both positively charged (in red), both negatively charged (in blue), and oppositely charged (in orange) hadrons in MC events with the relevant kinematical cuts (a) and additional asymmetric cuts on the momenta of the first hadron in the pair (b).}
\label{PLOT_SIV_QUARK_DIHAD_ID}
\end{figure}
 
  It is also interesting to study pairs with the same charge, as this will allow one to measure an asymmetry with the Sivers function convoluted with  dihadron fragmentation functions that are different from those for the oppositely charged hadrons (as expected in analogy to the results for the NJL-jet calculations of dihadron fragmentation functions that depend on $z_h$ and $M_h$ of Ref.~\cite{Matevosyan:2013aka}). In general, if we randomly assign labels $h_1$ and $h_2$ to same charged hadrons in the pair (and only apply symmetric momentum cuts), the R-type modulations are expected to vanish, since the difference of the transverse momenta of these hadrons in two similar events can have opposite signs and cancel any possible angular modulations. Nevertheless, we can generate non zero R-type SSAs for these pairs by ordering the hadrons in the pairs according to their energy fraction: the hadron $h_1$ would be the one with the larger value of $z$ (i.e. $z$ order the hadrons in a pair). Further, we  can enhance both R- and T-type modulations by imposing asymmetric cuts on the momenta of the hadrons in the pair. For example, setting the $z_1>z_{min}$ for the hadron labelled $h_1$ would preferentially select those hadrons produced at the initial stages of the  hadronization process for large enough $z_{min}$, while the average production rank of the hadron labelled $h_2$ would become larger  when $z_{min}$ is increased.  Further, we can learn about the relative rates of the different hadron pair production and their dependence on the struck quark's flavor from the plots in Fig.~\ref{PLOT_SIV_QUARK_DIHAD_ID}, that are the analogues of those for the single-hadron production in Fig.~\ref{PLOT_SIV_QUARK_ID}. Here we note that $h^+h^-$ pairs are produced at the highest rate for pairs with no additional cuts  shown in Fig.~\ref{PLOT_SIV_QUARK_DIHAD_ID}(a) and those with asymmetric cuts on the momenta of the hadrons in the pair shown in Fig.~\ref{PLOT_SIV_QUARK_DIHAD_ID}(b). The $h^+h^+$ pairs are produced primarily by the $u$ quark at a rate comparable to that of $h^+h^-$ pairs, while $h^-h^-$ pairs are produced almost equally by both $u$ and $\bar{u}$ quarks at a significantly smaller rate compared to $h^+h^-$ pairs. Moreover,  both the extracted asymmetries as functions of all the relevant variables and the  production rates (that affect statistical errors) for the two negatively charged hadrons are much smaller than for the other pairs, even with the additional asymmetric cuts on the momenta of the first hadron. Thus, we will present only the results for the $h^+h^+$ pairs.

\begin{figure}[tb]
\centering 
\subfigure[] {
\includegraphics[width=\ImS]{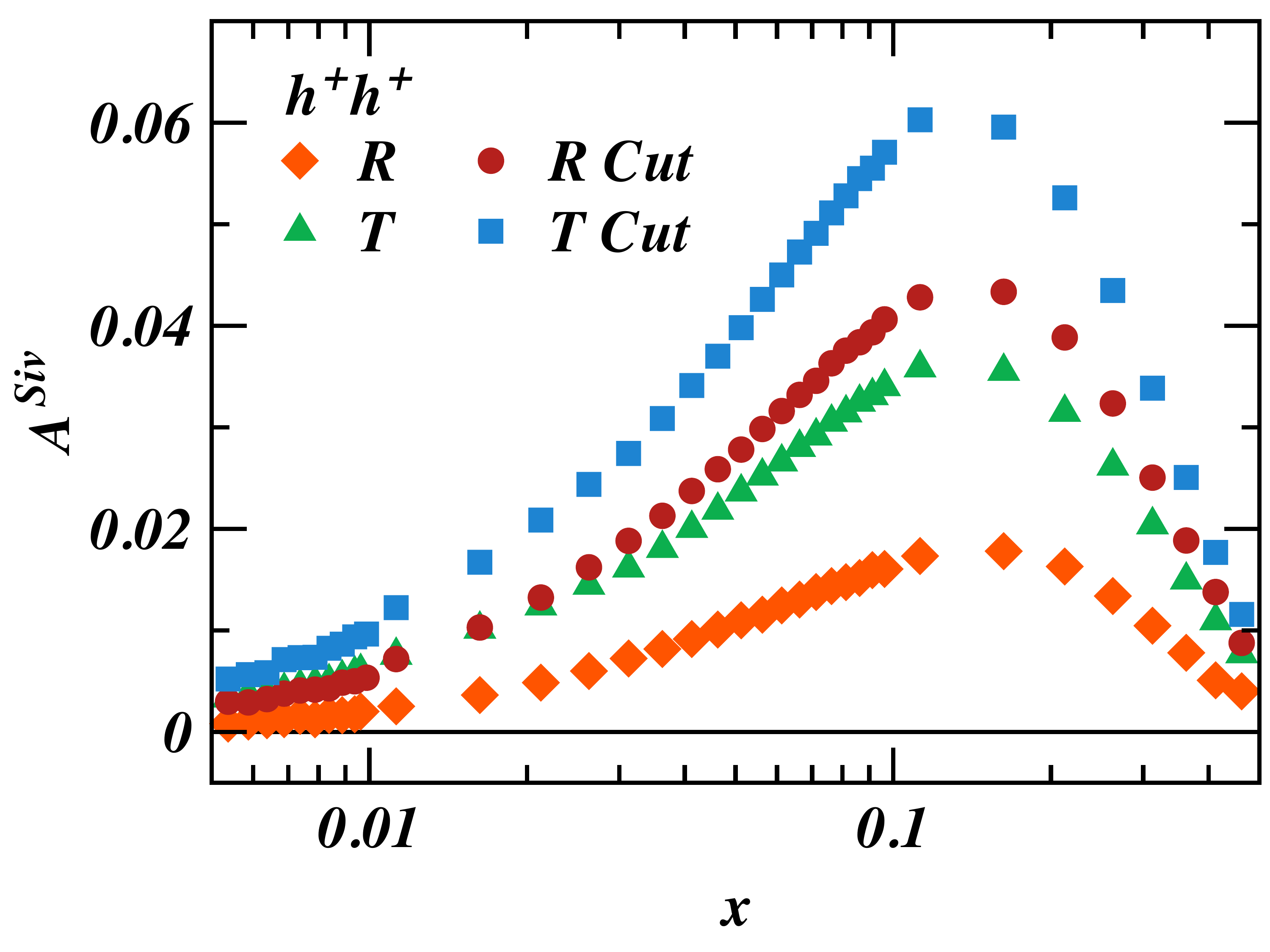}}
\\\vspace{-0.2cm}
\subfigure[] {
\includegraphics[width=\ImS]{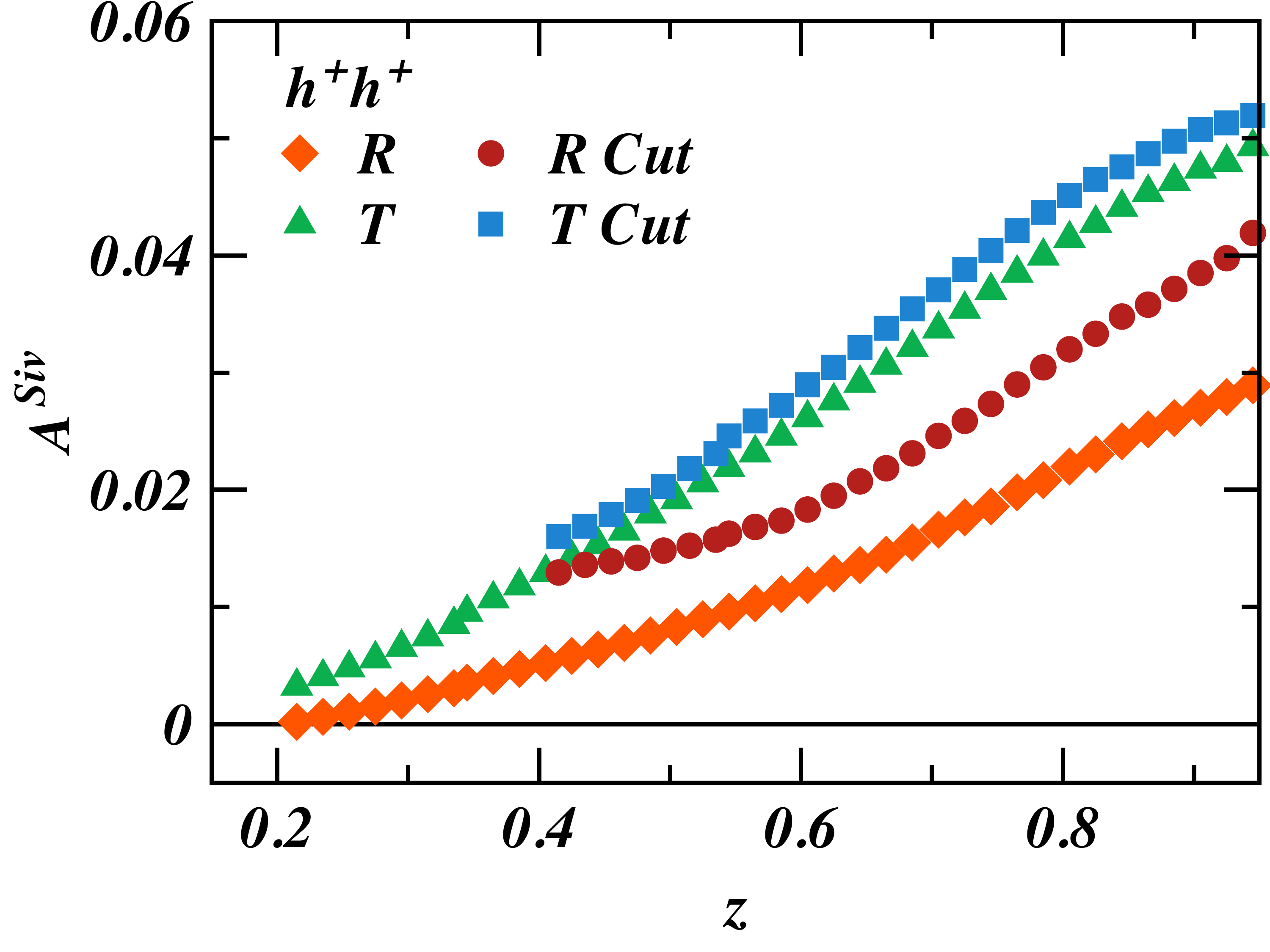}
}
\\\vspace{-0.2cm}
\caption{\MLEPT predictions for the dependence of the Sivers asymmetry on: (a) $x$, and (b) the total energy fraction $z$, in positively charged hadron pair production off a proton target for both $\fR$ and $\fT$ asymmetries integrated over $\vect{P}_T$ and $\vect{R}$, respectively. The bands labeled "Cut" are the results with the additional cut on the first hadron's momentum, as described in the text.}
\label{PLOT_SIV_2H_PLPL_X_Z}
\end{figure}

\begin{figure}[t]
\centering 
\subfigure[] {
\includegraphics[width=\ImS]{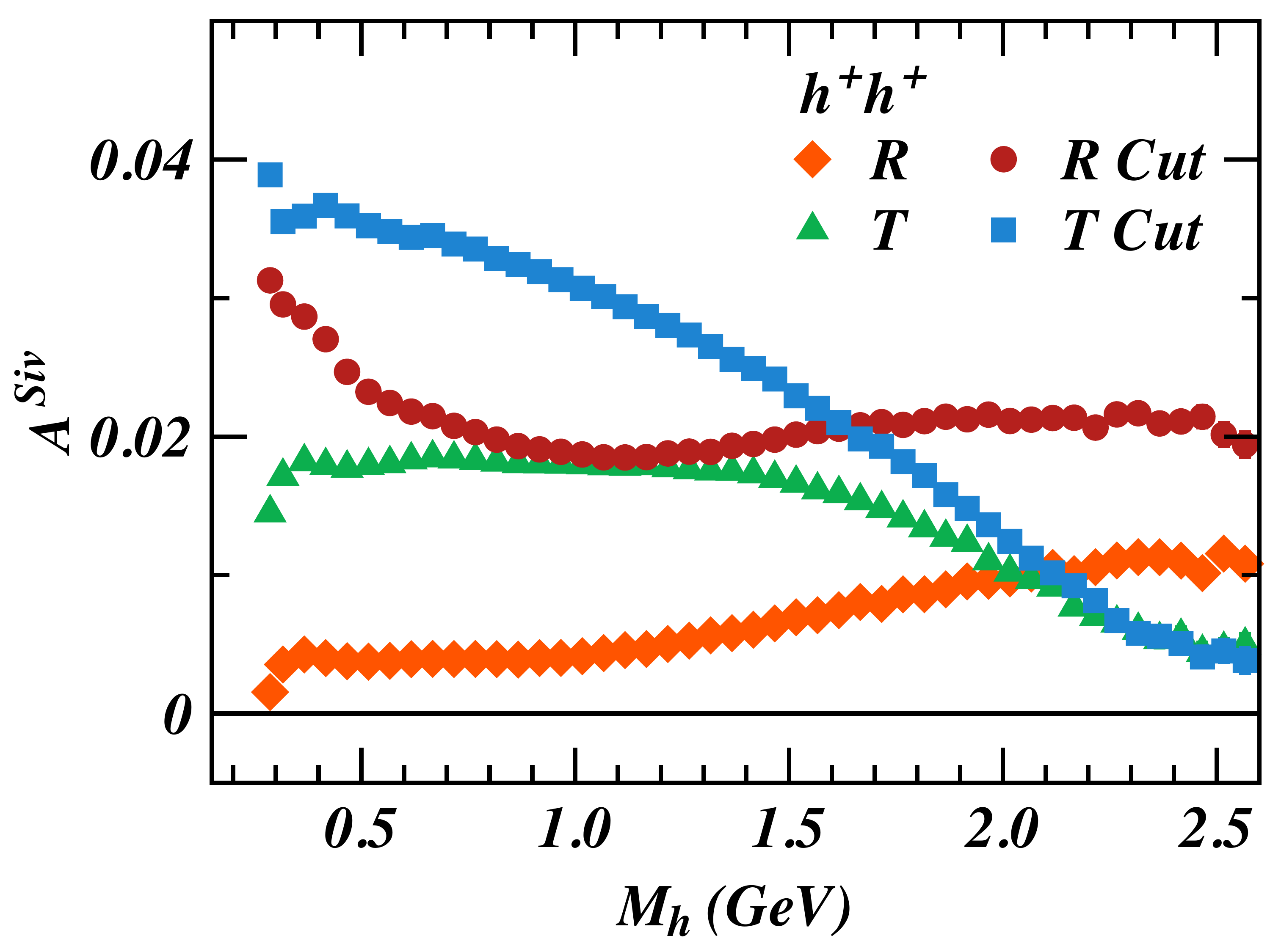}}
\\\vspace{-0.2cm}
\subfigure[] {
\includegraphics[width=\ImS]{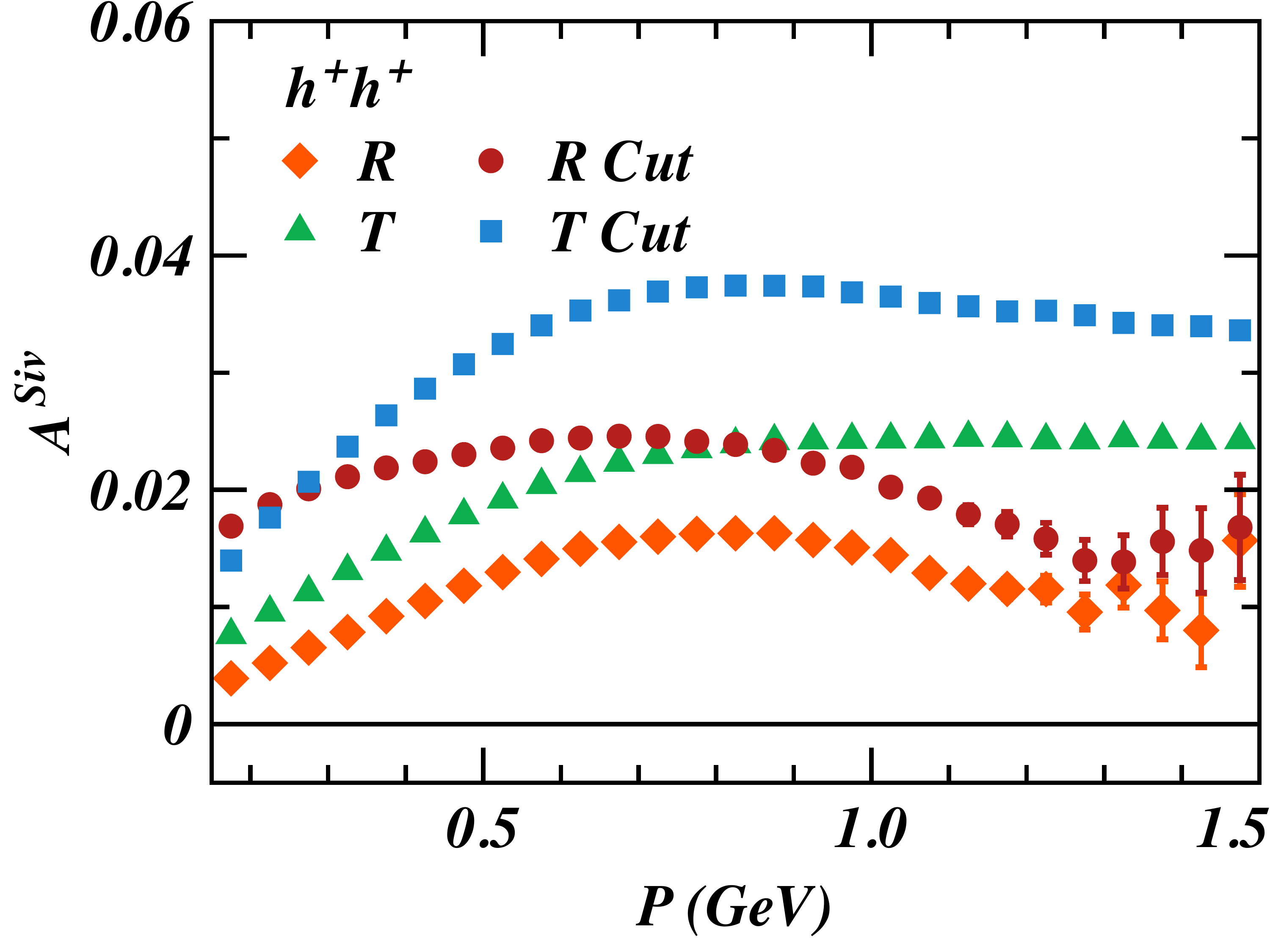}
}
\\\vspace{-0.2cm}
\caption{\MLEPT predictions for the dependence of the Sivers asymmetry on (a) $M_h$ and (b) the magnitude of the relevant momentum $P$, in positively charged hadron pair production off a proton target for both $\fR$ and $\fT$ asymmetries integrated over $\vect{P}_T$ and $\vect{R}$, respectively. The bands labeled "Cut" are the results with the additional cut on the first hadron's momentum, as described in the text.}
\label{PLOT_SIV_2H_PLPL_MH_PT}
\end{figure}

The plots in the Figs.~\ref{PLOT_SIV_2H_PLPL_X_Z} and \ref{PLOT_SIV_2H_PLPL_MH_PT} are the analogues of those in Figs.~\ref{PLOT_SIV_2H_MIPL_X_Z} and~\ref{PLOT_SIV_2H_MIPL_MH_PT} respectively, for the $h^+h^+$ pairs. We note that the asymmetries for the $h^+h^+$ pairs are enhanced when compared to the corresponding ones for $h^+h^-$ pairs. For example, the peak value of the $T\ Cut$ asymmetry shown as a function of $x$ is approximately $0.05$ for the $h^+h^-$ pairs in Fig.~\ref{PLOT_SIV_2H_MIPL_X_Z}(a), while for $h^+h^+$ pairs it is approximately $0.06$ in Fig.~\ref{PLOT_SIV_2H_PLPL_X_Z}(a). While this makes the asymmetries for $h^+h^+$ pairs easier to measure, \MLEPT simulations show that the number of $h^+h^+$ pairs produced within the kinematical limits employed in this work will be several times less than the $h^+h^-$ pairs, as shown in Fig.~\ref{PLOT_SIV_QUARK_DIHAD_ID}.

\section{Conclusions}
\label{SEC_CONC}

The experimental measurement of the Sivers effect in two-hadron semi-inclusive production in lepton-nucleon DIS was first proposed in Ref.~\cite{Kotzinian:2014lsa}, where only a brief summary of the results of parton model calculations of the relevant cross section terms were presented, along with \MLEPT predictions for the corresponding asymmetries for oppositely charged hadron pairs in COMPASS kinematics. This is the first article where we present the complete details of both phenomenological expressions for the relevant terms in the dihadron SIDIS cross section and independent numerical Monte Carlo predictions for the corresponding SSAs in the COMPASS kinematics. In this work we have expanded the calculations of the cross section terms relevant for the Sivers effect of \Eq{EQ_2H_SIG_SIV}, where the correlation between the transverse momentum of the unpolarized struck quark inside of the transversely polarized nucleon induces modulations~(\ref{EQ_2H_SIV_GEN}) with respect to the sines of the angles between each  hadron's  transverse momenta and the nucleon's transverse spin in the cross section of two-hadron SIDIS. We used simple parton model parametrizations of the unpolarized and Sivers PDFs, as well as the unpolarized DiFF of Eqs.~(\ref{EQ_UNPOL_PDF}-\ref{EQ_UNPOL_DIFF}) to perform the integrals over the transverse momenta in \Eq{EQ_2H_SIG_SIV}, with explicit expressions for the resulting cross section terms of \Eq{EQ_2H_XSEC_12RES} presented in the Appendix~\ref{SEC_APP_P12}. Further, in Appendix~\ref{SEC_APP_P12_1H} we calculated the explicit expression for the cross section of Eqs.~(\ref{EQ_2H_XSEC_INT1},\ref{EQ_2H_XSEC_INT2}), where one has integrated over the azimuthal angle of one of the hadrons in the pair. Here the cross section is still modulated with respect to the sine of the difference of the unintegrated hadron's azimuthal angle and $\fS$, with a coefficient involving the moments of the Fourier expansions of both the  unintegrated cross section terms $\sigma_{1}$ and $\sigma_{2}$. The cross section of two-hadron SIDIS is often expressed in terms of the vectors $\vect{P}_T$ and $\vect{R}$ which are the total and half of the relative transverse momenta of the hadron pair. We derived the corresponding relations in \Eq{EQ_2H_X_SEC_RT} for the cross section in Appendix~\ref{SEC_APP_RT}. Here again the cross section is modulated with respect to both $\sin(\fT-\fS)$ and $\sin(\fR-\fS)$, where the explicit expressions for the corresponding coefficients $\sigma_T$ and $\sigma_R$ in Eqs.~(\ref{EQ_APP_C_SIGMA_S}-\ref{EQ_APP_C_SIGMA_R}) show that both these terms are in general nonvanishing. Finally, in Appendix~\ref{SEC_APP_RT_1H} we present the explicit expressions for the cross section  \Eq{EQ_2H_X_SEC_RT}  integrated over  $\fR$ in \Eq{EQ_2H_X_SEC_INT_R} and over $\fT$ in \Eq{EQ_2H_X_SEC_INT_T}. Again, the integrated cross sections are modulated with respect to the sine of the unintegrated angle minus $\fS$, where the explicit expressions for the corresponding coefficients in Eqs.~(\ref{EQ_APP_D_SIGMA_U0},\ref{EQ_APP_D_SIGMA_T0},\ref{EQ_APP_D_SIGMA_T1},\ref{EQ_APP_D_SIGMA_R0},\ref{EQ_APP_D_SIGMA_R1}) prove that these terms are nonvanishing in general.

In the last part of Sec.~\ref{SEC_SIV_2H} we showed using leading order approximations, that relative transverse momentum $\vect{R}^P$ defined in Ref.~\cite{Bianconi:1999cd,Bacchetta:2003vn}, is independent of the transverse momentum of the fragmenting quarks' transverse momentum.  This entails that no associated Sivers effect should be observed for this choice of $\vect{R}^P$, as detailed in Ref.~\cite{Kotzinian:2014hoa}. Subsequently, experimental measurements of Sivers-type modulations involving both $\vect{R}$ (where we expect their presence) and $\vect{R}^P$ (where no modulations are expected) will serve as a test for applicability of the factorization framework and the leading order approximations used in this and earlier work.

 In Sec.~\ref{SEC_MLEPTO} we expanded the description of the \MLEPT MC event generator that incorporates both Sivers and Cahn effects. We have presented the results for (both positively and negatively charged) single-hadron Sivers asymmetries in COMPASS kinematics as functions of the Bjorken $x$, the hadron energy fraction $z$ and the transverse momentum $P_T$ in Fig.~\ref{PLOT_SIV_1H}. The results of the COMPASS measurements presented in the same figure indicate a good agreement of our MC results with the experiment. Further, we have presented the results for the dihadron Sivers asymmetries of oppositely charged hadron pairs in Figs.~\ref{PLOT_SIV_2H_MIPL_X_Z} and~\ref{PLOT_SIV_2H_MIPL_MH_PT} for both $\sin(\fT-\fS)$ and $\sin(\fR-\fS)$ modulations as functions of $x$, $z$, invariant mass $M_h$ and the relevant transverse momentum ($P_T$ or $R$). These results show nonzero modulations for all the considered dependencies, thus providing an {\it independent} proof that the Sivers effect  leads to both $\sin(\fT-\fS)$ and $\sin(\fR-\fS)$ modulations in the dihadron SIDIS cross section. Moreover, with an appropriate choice of kinematics such modulations are also present for the $h^+h^+$ hadron pairs, as shown in Figs.~\ref{PLOT_SIV_2H_PLPL_X_Z} and~\ref{PLOT_SIV_2H_PLPL_MH_PT}, with larger asymmetries than for the $h^+h^-$ pairs.

We conclude that the Sivers SSAs in the dihadron SIDIS process should be comparable to those for the single-hadron SIDIS. Thus, the experimental measurement of the Sivers SSAs for various hadron pairs  using the data collected at COMPASS and the future SIDIS experiments, such as those planned at JLAB12 and EIC, would contribute a large amount of information for extracting the Sivers PDFs. The unpolarized dihadron fragmentation functions of \Eq{EQ_2H_XSEC} can be studied in models, such as Refs.~\cite{Bacchetta:2006un,Casey:2012hg,Casey:2012ux,Matevosyan:2013aka}, and extracted using MC generators analogous to the studies  \cite{Courtoy:2012ry,Matevosyan:2013nla} of unpolarized DiFFs that depend on $z$ and $M_h$. Alternatively, these DiFFs and Sivers PDF can be extracted from experiment using our simple parametrizations of Sec.~\ref{SUB_SEC_XSEC_H1_H2} and the explicit cross section expressions presented in the Appendices, similar to the fits of the Sivers PDF in Ref.~\cite{Anselmino:2005nn}. In our future work we will use the versatile \MLEPT generator  along with state-of-the-art parametrizations of the Sivers PDFs to calculate projections for the Sivers SSAs in kinematics of various upcoming experiments, such as those proposed at JLAB12 and EIC. 

\FloatBarrier
\section*{Acknowledgements}

 We would like to thank Stefano Mellis for providing us with the parameters of the fits for the Sivers function for the sea quarks. A.K. was partially supported by CERN TH division and INFN Torino, and H.M. and A.T. were supported by the Australian Research Council through Grants  FL0992247 (AWT) and No.  CE110001004 (CoEPP) and by the University of Adelaide. 


\appendix

\begin{widetext}

\section{The Cross Section Terms Dervied Using the Parametric Forms for PDFs and DiFFs}
\label{SEC_APP_P12}

 In this section we present the explicit expressions for the cross section terms in \Eq{EQ_2H_XSEC_12RES} derived from \Eq{EQ_2H_SIG_SIV} using the parametrizations for PDFs and DiFFs of Eqs.~(\ref{EQ_UNPOL_PDF}-\ref{EQ_UNPOL_DIFF}),
along with the approximate relations for the transverse momenta of the produced hadrons of Eq.~(\ref{EQ_PT}).
The factor $C_0^{h_1 h_2}$ in the unpolarized term $\sigma_U$ in the \Eq{EQ_2H_XSEC_12RES} can be expressed as
\al
{
\label{EQ_APP_C0}
C_0^{h_1 h_2} 
=\frac{e^{-L}}{\pi^2 \lambda_{12}^4} 
\left( 1+c\ {(\PT{1}-z_1\vect{V}) \cdot (\PT{2}-z_2 \vect{V})} + {z_1  z_2 c}\ \frac{\mu_0^2 \nu_1^2 \nu_2^2}{\lambda_{12}^4} \right),
}
where
\al
{
\label{EQ_APP_A_NOTATION}
\lambda_{12}^4 &\equiv \nu_1^2 \nu_2^2 + \mu_0^2 (z_1^2 \nu_2^2 + z_2^2 \nu_1^2 ),
\\
L &\equiv   \frac{1}{\lambda_{12}^4}\left( \nu_2^2 {P}_{1T}^2 + \nu_1^2 {P}_{2T}^2 +\mu_0^2 (z_2 \vect{P}_{1T} - z_1 \PT{2})^2 \right),
\\
\vect{V} &\equiv \frac{\mu_0^2}{\lambda_{12}^4} (z_1 \nu_2^2 \PT{1}  + z_2 \nu_1^2 \PT{2})= \beta_1 \PT{1} + \beta_2 \PT{2},
\\
\beta_1 &\equiv \frac{z_1\mu_0^2 \nu_2^2}{\lambda_{12}^4}, 
\ 
\beta_2 \equiv \frac{z_2 \mu_0^2 \nu_1^2}{\lambda_{12}^4}.
}

The terms $C_{1,2}^{h_1 h_2}$ of the cross section in \Eq{EQ_2H_XSEC_12RES} induced by the Sivers effect, $\sigma_{1,2}$, can be expressed as
\al{
C_1^{h_1h_2} \equiv \frac{e^{-L_S}}{\pi^2 \lambda_{S12}^4}  (\beta_{S1}-c{B_1}),
 \\
 C_2^{h_1h_2} \equiv \frac{e^{-L_S}}{\pi^2 \lambda_{S12}^4}  (\beta_{S2}-c{B_2}),
}
where $L_S$, $V_S$, $\lambda_{S12}$, $\beta_{S1}$ and $\beta_{S2}$ are defined by replacing $\mu_0$ with $\mu_S$ in the corresponding expressions in~(\ref{EQ_APP_A_NOTATION})  for the unpolarized quantities and
\al
{
&
B_1 \equiv \frac{1}{2} \beta_{S2} \nu_2^2- \beta_{S1}(B_0+ z_1 z_2\frac{\mu_S^2  \nu_1^2 \nu_2^2}{\lambda_{S12}^4}),
\\
&
B_2 \equiv \frac{1}{2} \beta_{S1} \nu_1^2- \beta_{S2}(B_0+ z_1 z_2\frac{\mu_S^2  \nu_1^2 \nu_2^2}{\lambda_{S12}^4}),
}
\al
{
B_0 \equiv & 
\frac{z_1  z_2 \mu_S^2 \nu_1^2 \nu_2^2}{\lambda_{S12}^4}
-z_2\beta_{S1}(1-z_1\beta_{S1}) P_{1T}^2 - z_1\beta_{S2}(1-z_2\beta_{S2}) P_{2T}^2
\\ \non
&
+ ( 1-z_1\beta_{S1} - z_2\beta_{S2} + 2 z_1 z_2 \beta_{S1} \beta_{S2} ) P_{1T} P_{2T} \cos(\vf_1-\vf_2).
}
%

\section{Derivation of the Integrated Cross Section}
\label{SEC_APP_P12_1H}

In this section we derive the cross section (\ref{EQ_2H_SIG_SIV}) for two-hadron production in SIDIS process, integrated over the azimuthal angle of one of the hadrons. The dependence of the cross section on $\vf_1$ and $\vf_2$ is given by the explicit sine dependence of the Sivers terms (\ref{EQ_2H_SIV_GEN}) and by the dependence of $\sigma_{U\{,1,2\}}$ on $\cos(\vf_1-\vf_2)$. Thus, we first expand these structure functions as Fourier series in $\cos(\vf)$, with $\vf \equiv \vf_1-\vf_2$,
\al{
&
\sigma_i = \frac{1}{2\pi}\sum_{n=0}^\infty \sigma_{i,n} \cos(n\vf); 
\ \ 
i \in \{U,1,2\},
}
where we can access the $m$th moment using the orthogonality of the cosines
\al
{
&\non
\int_{-\pi}^\pi d\vf \cos(m\vf) \sigma_{i} = \frac{1}{2\pi} \sum_{n=0}^\infty \int_{-\pi}^\pi d\vf  \cos(m\vf) \cos(n\vf) \sigma_{i,n}
 = \frac{1+\delta_m^0}{2}\sigma_{i,m},
\\&
\sigma_{i,m} = \frac{2} {1+\delta_m^0}\int_{-\pi}^\pi d\vf \cos(m\vf) \ \sigma_i;
\ \ 
i \in \{U,1,2\}.
}

 Here we calculate the integral of the cross section over the azimuthal angle of the first hadron $\vf_1$
\al{
&
\frac{d \sigma^{ h_1 h_2 } }
{P_{1T}\ d P_{1T}\ d^2 \PT{2} }
=\int d\vf_1 C(x,Q^2) \left[ \sigma_{U} + S_T\left( \sigma_1 \frac{P_{1,T}}{M}\sin(\vf_1-\vf_S) +\sigma_2 \frac{P_{2,T}}{M}\sin(\vf_2-\vf_S)  \right)   \right].
}

In the unpolarized part, only the zeroth moment survives after the integration
\al{
\int d\vf_1 \sigma_{U} &
=  \sigma_{U,0} 
 = \sum_q e_q^2 f_1^q(x) D_{1q}^{h_1 h_2}(z_1, z_2) C_{0,0}^{h_1 h_2}.
}

Then, in the Sivers term involving $\sin(\vf_1-\vf_S)$, only the first moment remains after the integration
\al{
\int d\vf_1 \sigma_{1} \sin(\vf_1-\vf_S) &
= \frac{\sigma_{1,1} \sin(\vf_2-\vf_S) }{2}  
 = \frac{1}{2}\sum_q e_q^2 f_{1T}^{\perp q}(x) D_{1q}^{h_1 h_2}(z_1, z_2) C_{1,1}^{h_1 h_2} \sin(\vf_2-\vf_S).
}

Finally, for the Sivers term involving $\sin(\vf_2-\vf_S)$,
\al{
\int d\vf_1 \sigma_{2} \sin(\vf_2-\vf_S) &
= \sigma_{2,0} \sin(\vf_2-\vf_S) 
 = \sum_q e_q^2 f_{1T}^{\perp q}(x) D_{1q}^{h_1 h_2}(z_1, z_2) C_{2,0}^{h_1 h_2} \sin(\vf_2-\vf_S).
}

 Analogously, we can derive the expressions obtained after integrating over $\vf_2$. Then we can summarize  the results for integration over either $\vf_1$ or $\vf_2$,
\al{
&
\frac{d \sigma^{h_1h_2 } } {P_{1T}\ d P_{1T}\ d^2 \PT{2} } = C(x,Q^2)
\left[ 
\sigma_{U,0} +  S_T \left( \frac{P_{1T}}{2M} \sigma_{1,1}+\sigma_{2,0} \frac{P_{2T}}{M} \right) \sin(\vf_2-\vf_S)  \right],
\\&
\frac{d \sigma^{ h_1h_2 } }
{d^2 \PT{1}\ P_{2T}\ d P_{2T}}
= C(x, Q^2) 
\left[ 
\sigma_{U,0} +    S_T \left( \frac{P_{1T}}{M} \sigma_{1,0}+\sigma_{2,1} \frac{P_{2T}}{2M} \right) \sin(\vf_1-\vf_S)  \right].
}

Next  we calculate the zeroth moment of $C_0^{h^1h^2}$. We can easily see form \Eq{EQ_APP_C0}, that $C_0^{h_1h_2}$ depends linearly on $\cos(\vf)$, times an exponent in $\cos(\vf)$
\al{
C_{0}^{h_1 h_2} = E_0\  e^{- \xi \cos(\vf)} (c_0+ c_1 \cos(\vf)),
}
where
\al
{
&
E_0  \equiv \frac{1}{\pi^2 \lambda_{12}^4}  e^{-\left(\PT{1}^2 (\nu_2^2 + \mu_0^2 z_2^2) + \PT{2}^2 (\nu_1^2 + \mu_0^2 z_1^2) \right)/\lambda_{12}^4  },
\\&
\xi \equiv \frac{2 z_1 z_2 \mu_0^2 }{\lambda_{12}^4} P_{1T} P_{2T},
}
and
\al{
c_0 &= 1+
c
\left[ 
\frac{z_1  z_2 \mu_0^2 \nu_1^2 \nu_2^2}{\lambda_{12}^4}
-z_2\beta_1(1-z_1\beta_1)\PT{1}^2 - z_1\beta_2(1-z_2\beta_2)\PT{2}^2
\right]  ,
\\ 
c_1 &=  c\ ( 1-z_1\beta_1 - z_2\beta_2 + 2 z_1 z_2 \beta_1 \beta_2 ) P_{1T} P_{2T} \cos(\vf_1-\vf_2).
}

Then we  express the relevant moments of $C_{0}^{h_1 h_2}$ as
\al{
C_{0,0}^{h_1 h_2} &
= 2\pi E_0\ (c_0 I_0(\xi) + c_1 I_1(\xi)),
\\
C_{0,1}^{h_1 h_2} &
= 4\pi E_0 \left(c_0 I_1(\xi) + \frac{1}{2} c_1 \left[ I_0(\xi) + I_2(\xi) \right] \right),
}
where $I_n(a)$ are the modified Bessel functions of the first kind.

Next, we want to calculate the first two moments of $\sigma_1$, by first expressing it as
\al{
C_{1}^{h_1h_2}
= E_S e^{-\xi_S \cos(\vf)} (d_0 +d_1 \cos(\vf)),
}
where
\al
{
&
E_S  \equiv \frac{1}{\pi^2 \lambda_{S12}^4}  e^{-\left(\PT{1}^2 (\nu_2^2 + \mu_S^2 z_2^2) + \PT{2}^2 (\nu_1^2 + \mu_S^2 z_1^2) \right)/\lambda_{S12}^4  },
\\&
\xi_S \equiv \frac{2 z_1 z_2 \mu_S^2 }{\lambda_{S12}^4} P_{1T} P_{2T},
}
and
\al
{
d_0  & = \beta_{S1}  - c\ [  \frac{1}{2} \beta_{S2} \nu_2^2
- \beta_{S1}( 2 z_1 z_2\frac{\mu_S^2  \nu_1^2 \nu_2^2}{\lambda_{S12}^4}
-z_2\beta_{S1}(1-z_1\beta_{S1}) P_{1T}^2 - z_1\beta_{S2}(1-z_2\beta_{S2}) P_{2T}^2
) ],
\\
d_1  & =  c\ {\beta_{S1}}
( 1-z_1\beta_{S1} - z_2\beta_{S2} + 2 z_1 z_2 \beta_{S1} \beta_{S2} ) P_{1T} P_{2T} \cos(\vf_1-\vf_2).
}

Then we can easily calculate the first two moments of $C_{1}^{h_1 h_2}$
\al{
C_{1,0}^{h_1 h_2} &
= 2\pi E_S\ (d_0 I_0(\xi_S) + d_1 I_1(\xi)),
\\
C_{1,1}^{h_1 h_2} &
= 4\pi E_S \left(d_0 I_1(\xi_S) + \frac{1}{2} d_1 \left[ I_0(\xi_S) + I_2(\xi_S) \right] \right).
}

The expressions for the moments of $\sigma_2$ can be obtained by simply replacing the $1 \leftrightarrow 2$ indices in the corresponding expressions for the moments of $\sigma_1$.

\section{Derivation of the Cross Section in Terms of $\vT$ and $\vect{R}$ Variables}
\label{SEC_APP_RT}

 In this section we derive the formula for the cross section in terms of the transverse momenta $\vT$ and $\vect{R}$. It is easy to see from  Eq.~(\ref{EQ_T_R}) that the relevant variable change for the unpolarized term $\sigma_U$ in Eqs.~(\ref{EQ_2H_SIG_SIV},\ref{EQ_2H_XSEC_12RES}) only needs to be performed in the term $C_0^{h_1 h_2}$ from Eq.~(\ref{EQ_APP_C0})
\al{
C_0^{h_1 h_2}=
\frac{e^{-L}}{\pi^2 \lambda_{12}^4} 
&\left( 1+
c
\left[
 \frac{z_1  z_2 \mu_0^2 \nu_1^2 \nu_2^2}{\lambda_{12}^4}
 +(1-z_1(\beta_1 + \beta_2))(1-z_2(\beta_1 +  \beta_2))P_T^2/4 
\vphantom{\frac{1}{1}}
\right.\right. 
\\ \non& \left.\left. \vphantom{\frac{1}{1}}
 - (1-z_1(\beta_1 -\beta_2))(1+z_2(\beta_1 -\beta_2)) R^2
+ (z_1\beta_2 - z_2 \beta_1 + z_1z_2 (\beta_1^2-\beta_2^2) )  P_T R \cos(\vf)
\right] 
\right),
}
and
\al{
L &
= \frac{1}{\lambda_{12}^4}\left[
\frac{1}{4} P_T^2  (\nu_1^2 + \nu_2^2 + \mu_0^2(z_1-z_2)^2)
+ R^2  (\nu_1^2 + \nu_2^2 + \mu_0^2(z_1+z_2)^2)
+ \vT \cdot \vect{R} (\nu_2^2 -\nu_1^2 + \mu_0^2(z_2^2-z_1^2))
\right].
}
It is clear from the expression presented above, that  $\sigma_U$ is a function of $P_T, R$ and $\vT\cdot\vect{R}$ (or $\cos(\fT-\fR)$.
\\

Next, let's examine the $\sigma_{S}$ term in Eq.~(\ref{EQ_2H_SIV_GEN}), which we can rewrite as:
\al{
&\label{EQ_APP_C_SIGMA_S}
\sigma_{S} 
= \left(\sigma_1\frac{[\vect{S}_T,{\vT}/{2}  + \vect{R}]_3}{M} \right.
+
\left. \sigma_2 \frac{[\vect{S}_T, {\vT}/{2}  - \vect{R}]_3}{M} \right)
 = S_T  \left( \sigma_T \frac{P_T}{M} \sin(\fT-\fS) \right.
+
\left. \sigma_R \frac{R}{M} \sin(\fR-\fS) \right),
}
where
\al{
\sigma_T \equiv& \frac{\sigma_1 + \sigma_2}{2} = \sum\limits_q e_q^2 f_{1T}^{\perp q}(x)D_{1q}^{h_1h_2}(z_1,z_2)C_T^{h_1h_2},
\\ 
 \sigma_R \equiv& \sigma_1 - \sigma_2 = \sum\limits_q e_q^2 f_{1T}^{\perp q}(x)D_{1q}^{h_1h_2}(z_1,z_2) C_R^{h_1h_2}.
}
We can express $\sigma_T$ and $\sigma_R$ in terms of $\vT$ and $\vect{R}$ by calculating
\al
{\label{EQ_APP_C_SIGMA_T}
C_T^{h_1h_2}
=\frac{e^{-L_S}}{2\pi^2 \lambda_{S12}^4} \Bigg[ (\beta_{S1} + \beta_{S2}) \left(1+ z_1 z_2 c \frac{\mu_S^2  \nu_1^2 \nu_2^2}{\lambda_{S12}^4}\right)
-\frac{c}{2}( \beta_{S1} \nu_1^2 +  \beta_{S2} \nu_2^2 ) + c (\beta_{S1}+\beta_{S2})B_0 
\Bigg],
\\ \label{EQ_APP_C_SIGMA_R}
C_R^{h_1h_2} 
=\frac{e^{-L_S}}{\pi^2 \lambda_{S12}^4} \Bigg[ (\beta_{S1} - \beta_{S2}) \left(1+ z_1 z_2 c \frac{\mu_S^2  \nu_1^2 \nu_2^2}{\lambda_{S12}^4}\right)
-\frac{c}{2} (\beta_{S1} \nu_1^2 - \beta_{S2} \nu_2^2 ) + c (\beta_{S1}-\beta_{S2})B_0 
\Bigg].
}
where
\al{
B_0 =&  
\frac{z_1  z_2 \mu_S^2 \nu_1^2 \nu_2^2}{\lambda_{S12}^4}
+  ( 1-z_1(\beta_{S1}+ \beta_{S2})) ( 1-z_2(\beta_{S1}+ \beta_{S2})) \frac{P_T^2}{4}
\\ \non &
-   ( 1-z_1 (\beta_{S1}- \beta_{S2})) ( 1 + z_2 (\beta_{S1} -  \beta_{S2})) R^2
+ (z_1\beta_{S2}(1-z_2\beta_{S2}) -z_2\beta_{S1}(1-z_1\beta_{S1}) ) \vT \cdot \vect{R}.
}

It is easy to see that $C_R^{h_1h_2}$ is not vanishing when 
\al{
\beta_{S1} \neq \beta_{S2}
}
and/or 
\al{
\beta_{S1} \nu_1^2\neq \beta_{S2} \nu_2^2.
}
These conditions are satisfied, for example, for asymmetric ($z_1>z_2$) pairs.

\section{The Explicit Expressions of the Integrated Cross Section in Terms of $\vT$ and $\vect{R}$}
\label{SEC_APP_RT_1H}

In this section we derive the cross sections (\ref{EQ_2H_X_SEC_INT_T},\ref{EQ_2H_X_SEC_INT_R}) for two-hadron production in the SIDIS process, integrated over the azimuthal angle of the total or the relative transverse momenta of the hadron pair. The dependence on $\fR$ and $\fT$ of the cross section is given by the explicit sine dependence of the Sivers terms (\ref{EQ_2H_X_SEC_RT}), and by the dependence of $\sigma_{U\{,1,2\}}$ on $\cos(\fT-\fR)$. Here we can again employ the Fourier expansion of the relevant terms in the cross section
in
%
$\vf \equiv \fT -\fR$.
%

First we calculate the $\fR$ integrated results
\al{
\frac{d\sigma^{h_1h_2} }{d^2 \vT R d R} &=
\int d \fR \frac{d\sigma^{h_1 h_2}}{d^2\vT\, d^2\vect{R}\,} 
=C(x, Q^2)
\left[
 \sigma_{U,0} 
\vphantom{\frac{1}{1}} +
S_T \left(\frac{P_T}{M}\sigma_{T,0}+\frac{R}{2M}\sigma_{R,1} \right)\sin(\fT-\fS)
\right],
}

For the $\fT$ integrated results we have
\al{
\frac{d\sigma^{h_1h_2} }{P_T dP_T d^2\vect{R}} &=
\int d \fT \frac{d\sigma^{h_1 h_2}}{d^2\vT\, d^2\vect{R}\,}
=C(x, Q^2)
\left[
 \sigma_{U,0} 
\vphantom{\frac{1}{1}} +
S_T \left(\frac{P_T}{2M}\sigma_{T,1}+\frac{R}{M}\sigma_{R,0} \right)\sin(\vf_R-\fS)
\right].
}

We first calculate the zeroth moment of $\sigma_U$. This can be accomplished by first expressing ${C}_{0}^{h_1 h_2}$ in terms of the variables $\vT$ and $\vect{R}$ and introducing the following notation
\al
{
{C}_{0}^{h_1 h_2} = \tilde{E}_0 e^{- \tau \cos(\vf)} ( \tilde{c}_0 + \tilde{c}_1 \cos(\vf) ),
}
where
\al
{
&
\tilde{E}_0 \equiv  \frac{1}{\pi^2 \lambda_{12}^4}  e^{-\tilde{L}_0},
\\
&
\tilde{L}_0 \equiv \frac{1}{\lambda_{12}^4} [ 
\frac{1}{4} P_T^2  (\nu_1^2 + \nu_2^2 + \mu_0^2(z_1-z_2)^2)
+ R^2  (\nu_1^2 + \nu_2^2 + \mu_0^2(z_1+z_2)^2))
],
\\&
\tau \equiv \frac{\nu_2^2 -\nu_1^2 + \mu_0^2(z_2^2-z_1^2)}{\lambda_{12}^4} P_T R,
}
and
\al{
\tilde{c}_0 &\equiv 
 1+
c
\left[
 \frac{z_1  z_2 \mu_0^2 \nu_1^2 \nu_2^2}{\lambda_{12}^4}
 +
 ( (1-z_1(\beta_1 + \beta_2))(1-z_2(\beta_1 +  \beta_2))P_T^2/4 - (1-z_1(\beta_1 -\beta_2))(1+z_2(\beta_1 -\beta_2)) R^2 ) 
 \right],
\\
\tilde{c}_1 &\equiv 
c
\left[z_1\beta_2 - z_2 \beta_1 + z_1z_2 (\beta_1^2-\beta_2^2)\right] P_T  R.
}

 Then we can easily calculate 
%
\al
{
\label{EQ_APP_D_SIGMA_U0}
\sigma_{U,0} &
= 2 \pi  \sum_q e_q^2  f_{1}^{q}(x) D_{1q}^{h_1h_2}(z_1,z_2) \tilde{E}_0 ( \tilde{c}_0 I_0(\tau) +  \tilde{c}_1 I_1(\tau)).
}

Next, we calculate the zeroth and the first moments of $\sigma_T$, which are expressed in terms of the moments of ${C}_{T}^{h_1 h_2}$. To calculate the latter, first we introduce the following notation
\al
{
{C}_{T}^{h_1 h_2} = \tilde{E}_S e^{- \tau_S \cos(\vf)} ( \tilde{c}_{T0} + \tilde{c}_{T1} \cos(\vf) ),
}
where
\al
{
&
\tilde{E}_S \equiv \frac{1}{\pi^2 \lambda_{S12}^4}  e^{-\tilde{L}_S},
\\
&
\tilde{L}_S \equiv \frac{1}{\lambda_{S12}^4} [ 
\frac{1}{4} P_T^2  (\nu_1^2 + \nu_2^2 + \mu_S^2(z_1-z_2)^2)
+ R^2  (\nu_1^2 + \nu_2^2 + \mu_S^2(z_1+z_2)^2))
],
\\&
\tau_S \equiv \frac{\nu_2^2 -\nu_1^2 + \mu_S^2(z_2^2-z_1^2)}{\lambda_{S12}^4} P_T R,
}
and
\al
{
\tilde{c}_{T0}  \equiv \frac{c}{2}&
\left\{
 (\beta_{S1}+\beta_{S2})
 \left[ \frac{1}{c} + 2 z_1 z_2\frac{\mu_S^2  \nu_1^2 \nu_2^2}{\lambda_{S12}^4} +
 ( 1-z_1(\beta_{S1}+\beta_{S2}))(1 - z_2(\beta_{S1}+\beta_{S2}))  \frac{P_T^2}{4}
\right. \right.
\\ \non &
\left. \left.
-( 1-z_1(\beta_{S1}-\beta_{S2}))(1 + z_2(\beta_{S1}-\beta_{S2})) R^2
\vphantom{\frac{1}{1}}
\right]
-\frac{1}{2}( \beta_{S1} \nu_1^2 + \beta_{S2} \nu_2^2 )
\right\},
\\
\tilde{c}_{T1}  \equiv \frac{c}{2}&
 (\beta_{S1}+\beta_{S2})
  (z_1 \beta_{S2} - z_2 \beta_{S1} + z_1 z_2 (\beta_{S1}^2-\beta_{S2}^2) ) P_T R.
}
%


The calculation of the moments of $\sigma_T$ is then straightforward
\al
{
\label{EQ_APP_D_SIGMA_T0}
\sigma_{T,0} &
= 2 \pi  \sum_q e_q^2  f_{1T}^{\perp q}(x) D_{1q}^{h_1h_2}(z_1,z_2) \tilde{E}_S ( \tilde{c}_{T0} I_0(\tau_S) +  \tilde{c}_{T1} I_1(\tau_S)),
%
%
\\
\label{EQ_APP_D_SIGMA_T1}
\sigma_{T,1} &
= 4 \pi  \sum_q e_q^2 f_{1T}^{\perp q}(x) D_{1q}^{h_1h_2}(z_1,z_2) \tilde{E}_S \left( \tilde{c}_{T0} I_1(\tau_S) 
+ \frac{1}{2}[\tilde{c}_{T0} I_0(\tau_S) + \tilde{c}_{T1} I_1(\tau_S)]\right).
}

The calculation of the $\sigma_R$ moments is analogous to that for $\sigma_T$ case. First we introduce the notation
\al
{
C_R^{h_1h_2} &=
\tilde{E}_S e^{- \tau_S \cos(\vf)} ( \tilde{c}_{R0} + \tilde{c}_{R1} \cos(\vf) ),
}
where
\al
{
 \tilde{c}_{R0} \equiv c
 &
\left[
 (\beta_{S1} - \beta_{S2})
\left\{\frac{1}{c} + 2 z_1 z_2\frac{\mu_S^2  \nu_1^2 \nu_2^2}{\lambda_{S12}^4}
+ (1- z_1  (\beta_{S1}+\beta_{S2}))(1- z_2  (\beta_{S1}+\beta_{S2})) \frac{P_T^2}{4}
\right. \right.
\\& \non \left. \left. \vphantom{\frac{1}{1}}
-  (1- z_1  (\beta_{S1}-\beta_{S2}))(1+ z_2  (\beta_{S1}-\beta_{S2})) R^2
\right\}
-\frac{1}{2} (\beta_{S1} \nu_1^2 -\beta_{S2} \nu_2^2 ) 
\right],
\\
\tilde{c}_{R1} \equiv  c
&
  (\beta_{S1}-\beta_{S2}) ( z_1 \beta_{S2} - z_2 \beta_{S1} + z_1 z_2 (\beta_{S1}^2-\beta_{S2}^2) )  P_T R .
}

Then analogously to the case of $\sigma_T$, we  derive
\al{
\sigma_{R,0}&= 2 \pi  \sum_q e_q^2  f_{1T}^{\perp q}(x) D_{1q}^{h_1h_2}(z_1,z_2) \tilde{E}_S ( \tilde{c}_{R0} I_0(\tau_S) +  \tilde{c}_{R1} I_1(\tau_S)),
\label{EQ_APP_D_SIGMA_R0}
\\
\label{EQ_APP_D_SIGMA_R1}
\sigma_{R,1} &
= 4 \pi  \sum_q e_q^2 f_{1T}^{\perp q}(x) D_{1q}^{h_1h_2}(z_1,z_2) \tilde{E}_S \left( \tilde{c}_{R0} I_1(\tau_S) 
+ \frac{1}{2}[\tilde{c}_{R0} I_0(\tau_S) + \tilde{c}_{R1} I_1(\tau_S)]\right).
}

\end{widetext}

\bibliographystyle{apsrev}
\bibliography{fragment}

\end{document}